\documentclass[aps,pra,onecolumn,superscriptaddress, floatfix]{revtex4-2}
\usepackage{graphicx,bm}
\usepackage{amsmath}
\usepackage{physics}
\usepackage{amssymb}
\usepackage{upgreek}
\usepackage{url}
\usepackage{color}
\usepackage{siunitx}
\usepackage{setspace}
\usepackage{mathtools}
\usepackage{esvect}
\usepackage{caption}
\newcommand{\cvec}[1]{\vv{#1}}
\newcommand{\cmatrix}[1]{\hat{#1}}
\newcommand{\tmax}[0]{t_{\mathrm{max}}}
\newcommand{\set}[1]{\left\{#1\right\}}
\newcommand{\effective}[1]{#1_{\mathrm{eff}}}

\newcommand{\tcrit}[0]{t_{\mathrm{crit}}}
\newcommand{\eocs}[0]{emitter-occupied cavity state}

\begin{document}

\title{Optimising finite-time photon extraction from emitter-cavity systems}

\author{W. J. Hughes}
\email[email: ]{w.j.hughes@soton.ac.uk}
\affiliation{Optoelectronics Research Centre, University of Southampton, Southampton SO17 1BJ, UK}
\author{J. F. Goodwin}
\affiliation{Department of Physics, University of Oxford, Clarendon Laboratory, Parks Rd, Oxford, OX1 3PU, UK}
\author{P. Horak}
\affiliation{Optoelectronics Research Centre, University of Southampton, Southampton SO17 1BJ, UK}

\date{13 June 2024}

\begin{abstract}
We develop methods to find the limits to finite-time single photon extraction from emitter-cavity systems. We first establish analytic upper and lower bounds on the maximum extraction probability from a canonical $\Lambda$-system before developing a numeric method to optimise generic output probabilities from $\Lambda$-systems generalised to multiple ground states. We use these methods to study the limits to finite-time photon extraction and the wavepackets that satisfy them, finding that using an optimised wavepacket ranging between a sinusoidal and exponentially decaying profile can considerably reduce photon duration for a given extraction efficiency. We further optimise the rates of quantum protocols requiring emitter-photon correlation to obtain driving-independent conclusions about the effect of system parameters on success probability. We believe that these results and methods will provide valuable tools and insights for the development of cavity-based single photon sources combining high efficiency and high rate.
\end{abstract}
\maketitle

\section{Introduction}
\label{sec: introduction}

Single quantum emitters coupled to optical cavities have constituted a central platform for studying the interaction of light and matter~\cite{Hood:98, Birnbaum:05, Casabone:15}, and moreover present a possible implementation for many quantum information processes, including photon-photon gates~\cite{Stolz:22} and generating long-range interaction between matter-based qubits~\cite{Casabone:13, Haas:14, Ramette:22}. A broad class of applications, such as photonic information processing~\cite{Holleczek:16}, quantum networking~\cite{Wehner:18, Reiserer:22}, or networked modular quantum computation~\cite{Monroe:14}, could utilise emitter-cavity systems to produce single photons for the required protocols. The probability of photon extraction is typically of central importance, affecting the rate of heralded protocols~\cite{Cabrillo:99, Simon:03} and the fidelity of deterministic ones~\cite{Ritter:12}. 

There are several different approaches to produce single photons. To achieve fast extraction, one can directly excite the emitter~\cite{Bochmann:08}, but to approach the ultimate bound to photon extraction probability of $2C/(2C+1)$, determined solely by the cooperativity $C$~\cite{Goto:19, Law:97, Vasilev:10}, cavity-assisted Raman transitions~\cite{Maurer:04, Stute:12_2, Schupp:21}, or vSTIRAP~\cite{Kuhn:99} are commonly used. However, the photon production duration is typically much greater than direct excitation~\cite{Kuhn:02, Schupp:21} to reduce excited state population~\cite{Barros:09} or maintain adiabatic following~\cite{Hennrich:00} respectively. In quantum information applications, this increased photon duration leads not just to ultimately slower information processing, but increased decoherence of quantum information stored elsewhere during photon production~\cite{Monroe:13, Drmota:23, Tissot:24} and increased photonic losses in the longer fibre delay lines~\cite{Ward:22}. It is thus crucial to understand how to achieve both high efficiency and high rate photon extraction from emitter-cavity systems. 

In this work, we develop analytical and numerical approaches to find the limits set by system parameters on high-probability high-rate photon extraction, and how to saturate these limits. In Sec.~\ref{sec: three level system}, we develop upon previous approaches linking photon extraction probabilities to wavepacket shape~\cite{Vasilev:10, Utsugi:22} to analytically optimise these wavepackets and set upper and lower bounds on the maximum emission probability for a given production time. In Sec.~\ref{sec: numerical approach}, we build upon these ideas to develop a numeric procedure that can optimise a variety of probabilities, including the emission probability, for generalised versions of $\Lambda$-systems with multiple ground states. Finally, in Sec.~\ref{sec: results}, we use these tools to investigate the limits to finite-time photon extraction, and the photon wavepackets that satisfy them, before discussing protocols requiring emitter-photon correlations, taking remote entanglement generation~\cite{Luo:09} as a case study. Our approach of optimising the photon wavepacket directly is driving-independent, lending it complementary advantages to approaches that calculate dynamics directly from the driving pulse~\cite{Mucke:13, Ernst:23}, notably avoiding local optima in the selected driving parametrisation or ansatzes in the driving form.

Lastly, photon extraction and absorption, though not direct time reversals of each other, are linked by time reversal~\cite{Cirac:97}, with the efficiency of extraction matching that of absorption for time-reversed control drives and wavepacket profiles~\cite{Gorshkov:07_2}. Our results for photon extraction will therefore correspond to analogous results for photon absorption, on which there is extensive literature~\cite{Gorshkov:07_1, Gorshkov:07_2, Dilley:12, Giannelli:18}, but to avoid confusion, we will discuss only photon extraction contexts.

\section{Three level $\Lambda$-system}
\label{sec: three level system}
\subsection{Model}
\label{subsec: three level model}
\begin{figure}
\centering 
\captionsetup{width=0.95\textwidth}
\includegraphics*[width=0.56\linewidth]{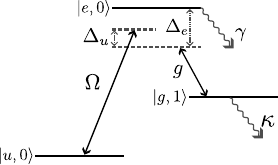}
\caption[Three-level system diagram]{The level scheme for the $\Lambda$-emitter coupled to a single cavity mode. The Hilbert space contains three levels that are all tensor products of the emitter's electronic state and the cavity mode Fock state. These levels are the initial emitter state with no cavity photon $\ket{u,0}$, an excited emitter state with no cavity photon $\ket{e,0}$, and a final emitter state with a cavity photon $\ket{g,1}$. The excited state and final state are coupled by the cavity coupling $g$, with the excited state detuned from resonance with the cavity by $\Delta_e$. The initial and excited state are coupled by a laser field with complex time-dependent coupling $\Omega$. The detuning of the laser field from Raman resonance with the cavity is $\Delta_u$, although this is only nominal as the laser field can be adjusted to include an arbitrary detuning in its time dependence. The excited state amplitude decays at a rate $\gamma$ due to spontaneous emission, and the cavity state amplitude at a rate $\kappa$ due to field loss from the cavity.} 
\label{fig: three level system}
\end{figure}

The model we use for the emitter-cavity photon-generation system is the canonical $\Lambda$-type emitter coupled to a single cavity mode. The emitter level structure contains two ground states, $\ket{u}$ and $\ket{g}$, which are both coupled by electromagnetic transitions to an excited state $\ket{e}$ which decays via spontaneous emission to any mode except the single cavity mode with amplitude decay rate $\gamma$. The emitter couples to a single cavity mode, whose Hilbert space contains only the vacuum state $\ket{0}$ and the single photon state $\ket{1}$. The cavity field amplitude decay rate is $\kappa$. In real systems, this decay rate comprises a `useful' decay rate through the desired partially-transmissive mirror $\kappa_T$, and a `parasitic' decay rate $\kappa_I$ which includes other losses such as scattering, absorption, and diffraction. To reduce the number of variables, we assume $\kappa_I=0$ and therefore $\kappa_T = \kappa$. However, the coherent dynamics of the system are determined solely by $\kappa$, therefore all future results will hold for cavities with $\kappa_I \neq 0$ provided any output probabilities are attenuated by $\kappa_T / (\kappa_T + \kappa_I)$. The coupling between the excited atomic state with an empty cavity $\ket{e,0}$, and the ground state with an occupied cavity $\ket{g,1}$ is, by convention, $g$, with detuning $\Delta_e$ of the $\ket{e}\rightarrow\ket{g}$ emitter transition frequency from the cavity mode frequency. A diagram of the system is depicted in Fig.~\ref{fig: three level system}.

In a general photon production process, the cavity is initially vacant, the emitter is prepared in state $\ket{u}$, and a time-dependent driving pulse $\Omega(t)$ is applied to transfer population, via $\ket{e,0}$, to $\ket{g,1}$. The single photon then leaks out of the cavity and is collected.  This $\Lambda$-system model is sufficiently general to describe photon production through direct excitation, cavity-assisted Raman transition or vSTIRAP through appropriate choices of the driving pulse $\Omega(t)$ and detuning $\Delta_e$~\cite{Goto:19}. Regardless of the driving method chosen, the maximum photonic output of the system is 
\begin{equation}
    \begin{aligned}
        P_{\kappa}^{(a)} & = \frac{2C}{2C+1}, \\
        C & = \frac{g^2}{2\kappa\gamma},
    \end{aligned}
    \label{eq: Lambda optimum output}
\end{equation}
where the dimensionless $C$ is the cooperativity~\cite{Goto:19}. This performance is achieved in the adiabatic (i.e. infinite extraction time) limit, and the goal of this manuscript is to determine similar limits for non-adiabatic timescales.

The coherent Hamiltonian for this model is
\begin{equation}
H = \Delta_u \dyad{u,0} + \Delta_e \dyad{e,0}  + \left[-i\Omega^* \dyad{u,0}{e,0} + \mathrm{h.c.}\right] + \left[ig \dyad{e,0}{g,1} + \mathrm{h.c.}\right], 
\label{eq: three level Hamiltonian}
\end{equation} 
where the factor of the imaginary unit $i$ before $g$ is made for algebraic convenience. The nominal detuning of the initial state from Raman resonance with $\ket{g,1}$ is $\Delta_u$, which is usually set to zero in monochromatic driving schemes. 

The system dynamics also feature two incoherent decay channels, spontaneous emission from the excited state and decay of the cavity mode, which may be included in the model by using the master equation. However, for emitter-cavity systems operating as photon sources, the dynamics are only relevant if the cavity emits a photon. This means that, on the assumption that no decay process can be followed by subsequent cavity decay, a simpler non-Hermitian Hamiltonian
\begin{equation}
H^{\mathrm{NH}} = H -i\gamma \dyad{e,0} - i\kappa \dyad{g,1},
\label{eq: non hermitian hamiltonian}
\end{equation}
can be used. The assumption that no decay process is followed by cavity decay is typically valid for cavity decay itself (provided the state $\ket{g}$ is stable on the timescale of the photon production process). However, spontaneous emission followed by cavity decay can also occur. This causes the emission of a probabilistic mixture of photon wavepackets, known as temporal mixing, which reduces the coherence and indistinguishability of photon wavepackets~\cite{Fischer:17}, resulting in major fidelity reductions for certain quantum protocols~\cite{Meraner:20, Krutyanskiy:23}. Choosing a $\Lambda$-system where the excited state decay has a low branching ratio to the initial state $\ket{u}$ strongly mitigates this effect~\cite{Walker:20}. Because temporal mixing errors can be so destructive to the coherence of the output wavepacket, we assume that the $\Lambda$-system has been chosen such that temporal mixing is negligible and, therefore, that the non-Hermitian approach is applicable. In the case that temporal mixing is not negligible, the results we derive maximise the probability of an emission event that is not preceded by a spontaneous emission event.

Finally, we note that recent work~\cite{Kikura:24} has proposed the use of multiple excited states to realise the reduced temporal mixing found in systems with low branching ratios to $\ket{u}$, but in systems that do not naturally have that structure. At the end of Sec.~\ref{sec: bounded approach}, we discuss how our results also apply to these systems.

\subsection{Bounded optimisation approach}
\label{sec: bounded approach}
Using the non-Hermitian approach, our analysis initially follows Goto (2019) \cite{Goto:19} to link the probabilities of cavity emission and spontaneous emission to the photon shape, and then, as suggested in Vasilev (2010) \cite{Vasilev:10}, optimises this shape to yield the maximum output.

We begin by expressing explicitly the non-Hermitian Hamiltonian Eq.~(\ref{eq: non hermitian hamiltonian})

\begin{equation}
\begin{aligned}
\ket{\Psi(t)} & = \alpha_u \ket{u,0} + \alpha_e \ket{e,0} + \alpha_g\ket{g,1}, \\
\dot{\alpha_u} & = -i\Delta_u\alpha_u - \Omega^*\alpha_e, \\
\dot{\alpha_e} & = -(\gamma+i\Delta_e)\alpha_e+\Omega\alpha_u+g\alpha_g, \\
\dot{\alpha_g} & = -\kappa\alpha_g-g\alpha_e ,
\end{aligned}
\label{simple lambda system equations}
\end{equation}
where $\ket{\Psi(t)}$ is the wavefunction of the emitter-cavity system. Rearrangement of the last relation yields 
\begin{equation}
\alpha_e = -\frac{1}{g}\left(\kappa \alpha_g + \dot{\alpha_g} \right),
\label{eq: excited state wavefunction}
\end{equation}
which expresses the wavefunction component of $\ket{e,0}$ through the component in the state $\ket{g,1}$. 

Throughout the photon generation process, the three level system with two decay channels has probabilities of occupation split across five categories: $P_u(t)$, $P_e(t)$, and $P_g(t)$ are the occupation probabilities of states $\ket{u,0}$, $\ket{e,0}$, and $\ket{g,1}$ at time $t$ respectively, $P_{\gamma}(t)$ is the probability that there has been decay by spontaneous emission by time $t$, and $P_{\kappa}(t)$ is the probability that there has been photonic decay from the cavity by time $t$.

These probabilities, with the exception of $P_u$, can be written using Hermitian integrals of the form
\begin{equation}
I_{(nm)}(t) = \int_0^t \left(\overset{n\cdot}{\alpha_g}\overset{m\cdot}{\alpha_g}^* + \overset{m\cdot}{\alpha_g}\overset{n\cdot}{\alpha_g}^*\right)\, dt,
\label{eq: integral notation definition}
\end{equation}
where $n\cdot$ is a shorthand for $n$ copies of $\cdot$ to indicate $n$ time derivatives. Using this notation, the probabilities are

\begin{equation}
\begin{aligned}
P_{\kappa}(t) &= \kappa I_{(00)}(t), \\
P_g(t) &= I_{(10)}(t), \\
P_{\gamma}(t) &= \frac{\gamma\kappa^2}{g^2}I_{(00)}(t) + \frac{2\gamma\kappa}{g^2}I_{(10)}(t) + \frac{\gamma}{g^2}I_{(11)}(t), \\
P_{e}(t) &= \frac{\kappa^2}{g^2}I_{(10)}(t) + \frac{\kappa}{g^2}I_{(20)}(t) + \frac{\kappa}{g^2}I_{(11)}(t) + \frac{1}{g^2}I_{(21)}(t) + \frac{1}{g^2}\abs{\dot{\alpha_g}(0)}^2, \\
P_{u}(t) &= 1 - P_{\kappa}(t) - P_g(t) -  P_{\gamma}(t) - P_{e}(t), 
\end{aligned}
\label{eq: simple lambda probabilities}
\end{equation}
where the derivations are given in App.~\ref{app: calculating probability functions}.

Assuming that any driving pulse $\Omega(t)$ can be applied, the $\alpha_e(t)$ required to produce a desired $\alpha_g(t)$ is possible provided there is sufficient probability $P_u(t)$ at all times to control the dynamics. In particular, an unphysical wavepacket $\alpha_g(t)$ will cause $P_u(t)$, which is calculated through probability conservation, to drop below zero at some time during the process. Therefore, a physical solution should satisfy $P_u(t) \geq 0, \, \, 0 \leq t \leq T$, with the optimal solution having $P_u(\tmax)=0$ for some time $\tmax$ during the process (see~\cite{Vasilev:10, Utsugi:22} for previous uses of this condition). This condition is troublesome to impose analytically as it applies a separate constraint for every time during photon production. Instead, we demand that no probability remains in the initial or excited state at the end of the process ($P_u(T)+P_e(T)=0$), which can be simply enforced. This `upper bound constraint' is less restrictive than the true constraint in the sense that any dynamics satisfying the true constraint has output probability less than or equal to some dynamics satisfying the upper bound constraint. This can be seen by noting that a solution satisfying the true constraint must satisfy $P_u(T) + P_e(T) \geq 0$, where the inequality can be converted to the equality of the upper bound constraint by uniformly increasing the scale of $\alpha_g$, and therefore increasing the output probability. Solutions derived under the upper bound constraints therefore constitute upper bounds to the optimum photon extraction probabilities, and will be denoted with a superscript $(U)$.

The constraint $P_u(T)+P_e(T)=0$ can be rewritten
\begin{equation}
1 = P_{\kappa}^{(U)}(T) + P_g^{(U)}(T) +  P_{\gamma}^{(U)}(T),
\label{eq: upper bound probability constraint}
\end{equation}
as the five probabilities must always sum to unity. This yields a  final output probability
\begin{equation}
\begin{aligned}
P_{\kappa}^{(U)}(T) & = \frac{1}{1+m}, \\
m & = \frac{P_g^{(U)}(T) +  P_{\gamma}^{(U)}(T)}{P_{\kappa}^{(U)}(T)} \equiv \frac{F}{G},
\label{eq: single lambda minimisation setup}
\end{aligned}
\end{equation}
where $m$ is a scalar determined by the shape of the photon wavepacket and $F$ and $G$ are functionals of $\alpha_g$ introduced for notational convenience.

As seen from Eq.~(\ref{eq: single lambda minimisation setup}), the optimum output probability occurs when $m$ is minimised. To minimise $m$, assume that $\alpha_g$ changes by $\delta \alpha_g$, causing a change in $F$ ($G$) of $\delta F$ ($\delta G$). The condition that $m$ is an extremum gives $\delta F /\delta G = F/G$. Now define a shape functional

\begin{equation}
S_q = P_g^{(U)}(T) +  P_{\gamma}^{(U)}(T) - q P_{\kappa}^{(U)}(T) = F-qG,
\label{S_q definition}
\end{equation}
for a general scalar $q$. Extrema of $S_q$ satisfy $\delta F /\delta G = q$. Therefore, the procedure to find extrema of $m$ with respect to changes in the photon shape $\alpha_g$ is a two-stage process: first, find the extrema of $S_q$ for all $q$ and second, check that $q=F/G=m$. In other words, stationary solutions must satisfy
\begin{equation}
    \begin{aligned}
        \delta S_q & = 0, \\
        q & = \frac{F}{G} \equiv m,
    \end{aligned}
    \label{eq: single lambda solution conditions}
\end{equation}
simultaneously, and the optimum solution is that with the smallest $q$. To satisfy the first condition of Eq.~(\ref{eq: single lambda solution conditions}), we find stationary $S_q$ according to 
\begin{equation}
\begin{aligned}
S_q & = \int_{0}^{T} L_q \, dt, \\
L_q & = \left(\frac{\gamma\kappa^2}{g^2} - q\kappa\right) \alpha_g \alpha_g^* + \left(1 + \frac{2\gamma\kappa}{g^2}\right) \dot{\alpha_g} \alpha_g^*  + \frac{\gamma}{g^2} \dot{\alpha_g} \dot{\alpha_g}^* + \mathrm{c.c.},
\end{aligned}
\end{equation}
where c.c. is a shorthand for complex conjugate, and the `effective Lagrangian' $L_q$ has been written explicitly using Eq.~(\ref{eq: simple lambda probabilities}). Using the Euler-Lagrange equations to minimise this functional
\begin{equation}
\frac{\partial L_q}{\partial \alpha_g} = \frac{d}{dt}\left(\frac{\partial L_q}{\partial \dot{\alpha_g}}\right),
\end{equation}
produces a differential equation
\begin{equation}
\ddot{\alpha_g} + \kappa^2 \left(2 C q -1\right) \alpha_g = 0,
\label{eq: analytic alpha g eom}
\end{equation}
for the wavefunction of the cavity mode. This equation has two types of solution. If $q < 1/(2C)$, the solutions are hyperbolic functions, whereas in the opposite case, they are trigonometric functions. It is shown in App.~\ref{app: upper bound does not have hyperbolic solution} (and could also be inferred directly from Eq.~(\ref{eq: Lambda optimum output})) that the hyperbolic solution produces $m \geq 1/(2C)$, and therefore can never satisfy the second constraint of Eq.~(\ref{eq: single lambda solution conditions}).
This means that the required $q\geq 1/(2C)$, resulting in trigonometric solutions. Setting the boundary condition that the cavity is vacant at the start of the process gives 

\begin{equation}
\begin{aligned}
\alpha_g(t) & = A \sin (\omega_q t), \\
\omega_q^2 & = \kappa^2 \left(2 C q -1\right),
\end{aligned}
\label{alpha g general solution}
\end{equation}
where $A$ is a free parameter that can be set to adjust the sum of probabilities. The probabilities from Eq.~(\ref{eq: simple lambda probabilities}) can be calculated explicitly using this solution: 

\begin{equation}
\begin{aligned}
P_{\kappa}(t) &= A^2 \left(\kappa t - \frac{\kappa}{2 \omega_q}\sin(2\omega_q t)\right), \\
P_g(t) &= A^{2} \sin^{2}{\left(\omega_{q} t \right)}, \\
P_{\gamma}(t) &= \frac{P_{\kappa}(t)}{2C} + A^2 \left\{\frac{1}{C}\sin^2(\omega_q t) + \frac{\omega_q^2}{\kappa^2} \frac{1}{2C}\left(\kappa t + \frac{\kappa}{2 \omega_q}\sin(2\omega_q t)\right)\right\}, \\
P_{e}(t) &= A^2 \left(\frac{\omega_q}{g}\cos(\omega_q t)+ \frac{\kappa}{g}\sin(\omega_q t)\right)^2.
\end{aligned}
\label{eq: upper bound probabilities}
\end{equation}
Substituting these results at time $T$ into  the second condition of Eq.~(\ref{eq: single lambda solution conditions}) ($q=F/G$) results in  
\begin{equation}
\cos(2\omega_q T)-\frac{2\omega_q}{2\kappa(1+C)}\sin(2\omega_q T)=1.
\label{omega_m restriction}
\end{equation}
This equation has many solutions, however, when satisfied, $q = F/G=m$, and therefore, to minimise $m$, the solution of interest has the minimum $q$ and therefore $\omega_q$. There is a solution at $\omega_q = \pi/T$, corresponding to a cavity wavefunction $\alpha_g$ (and thus photon wavepacket) of sinusoidal amplitude, beginning at zero and first returning to zero again at time $T$. However, there is a better solution at a lower $q$, which retains a sinusoidal shape, but with a slower temporal frequency $\omega_q$ so that the photon does not fully close by time $T$.

Thus the upper bound can be summarised
\begin{equation}
\begin{aligned}
1 & = \cos(2\omega_m T)-\frac{2\omega_m}{2\kappa(1+C)}\sin(2\omega_m T), \\
m & = \frac{1}{2C}\left(\left(\frac{\omega_m}{\kappa}\right)^2 + 1\right), \\
P_{\kappa}^u & = \frac{1}{1+ m}, \\
\end{aligned}
\label{eq: single lambda output final result}
\end{equation}
where $m$ is taken as the smallest value to satisfy the first equation, which must be solved numerically. The same optimisation could have been performed using Lagrange multipliers, but in that approach it is less clear which stationary solution is the global minimum and it is not emphasised so strongly that the optimisation performed optimises the photon shape.

The upper bound is an output optimised under the condition that the population remaining in the cavity $P_g$, emitted spontaneously $P_{\gamma}$ and decayed through the cavity $P_\kappa$ sum to unity at time $T$. This is less restrictive than the true condition that these probabilities, in addition to the probability in the excited state $P_e$ reaches unity at some point in the process. We obtain a lower bound to the output probability $P_{\kappa}^l$ by assuming the photon retains the same amplitude profile, but reduced in scale to satisfy the true probability constraint. This is a lower bound because the photon shape has not been optimised for the true constraint.

A potential objection against the validity of these sinusoidal solutions is that the initial occupation of the excited state $\ket{e,0}$ is non-zero, whereas the problem stipulates that the system is prepared in state $\ket{u,0}$. However, provided arbitrary driving is possible, probability can be transferred infinitely quickly from $\ket{u,0}$ to $\ket{e,0}$ at the beginning of the process. This means that the lower bound photon solution \textit{can} be produced from an initial state of $\ket{u,0}$ in a time only infinitesimally longer that $T$, and therefore the bounds apply unchanged in the case that the excited state is initially vacant. Indeed the ability to set a non-zero occupation of $\ket{e,0}$ at time $t=0$ is advantageous because it means that the effect of arbitrarily strong driving at the initial time can be captured without requiring these troublesome dynamics be modelled explicitly. 

Finally, it is worth noting that, while the bounds have been derived for a $\Lambda$-system, they are applicable to a wider class of systems. This is because the level structure through which wavefunction amplitude is delivered to the excited state is not relevant provided it does not restrict the possible $\alpha_e(t)$. In particular, as mentioned in Sec.~\ref{sec: three level system}, a recent paper~\cite{Kikura:24} has suggested the use of two excited states to reduce photon indistinguishability due to temporal mixing. Provided both of the driving fields used in that scenario can realise arbitrary amplitude profiles such that the occupation of the additional excited state is always negligible, a very similar derivation to the above can be made for those systems. This finds identical bounds to the $\Lambda$ system, which extends to finite time the equivalent adiabatic result described in that paper.

The upper and lower bounds described in this section can be readily calculated, but before those results are presented in Sec.~\ref{sec: results}, the numerical method will be developed.

\section{Numerical Method for Generalised $\Lambda$-systems}
\label{sec: numerical approach}
\subsection{Defining the system}
The results presented in Sec.~\ref{sec: bounded approach} set useful bounds on the limits of performance for $\Lambda$-systems, but there are benefits to a more flexible numeric approach. Firstly, while we know that the true limit to performance for the $\Lambda$-system lies between the upper and lower bounds found in Sec.~\ref{sec: bounded approach}, we do not know where between these bounds the limit lies. Secondly, real emitter level structures often contain additional decay channels near-resonant with the cavity modes, which, in atom or ion emitters, would typically be to alternative sublevels within the fine or hyperfine structure. Often, these additional levels must be included to make simulations consistent with experiment~\cite{Mucke:13, Barrett:18, Ernst:23}, and thus would ideally feature in our model. Finally, this extra structure can also be utilised to perform protocols which produce single photons entangled with their emitter~\cite{Simon:03, Wilk:07}, meaning that the ability to model systems with additional ground states is essential to determine the limits of these protocols.

\begin{figure}
\centering 
\captionsetup{width=0.95\textwidth}
\includegraphics*[width=0.56\linewidth]{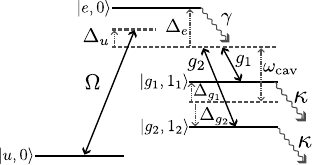}
\caption[Multi-level system diagram]{The level scheme for the generalised $\Lambda$-system coupled to a cavity mode, here depicted with two \eocs s. Compared to Fig.~\ref{fig: three level system}, the Hilbert space of the system now contains an arbitrary number of \eocs s (indexed by $j$) with cavity mode occupation $\ket{g_j,1_j}$ coupled to the excited state with respective coupling $g_j$. The \eocs s are each detuned from a nominal central energy level by detuning $\Delta_{g_j}$ from which the excited state is itself detuned by $\Delta_e$ when including the cavity photon energy $\omega_{\mathrm{cav}}$.} 
\label{fig: multi level system}
\end{figure}

The ideas presented in Sec.~\ref{sec: bounded approach} inspire a numerical method presenting these benefits. The systems modelled again contain a single initial state $\ket{u,0}$ and excited state $\ket{e,0}$, but now potentially multiple distinct states with occupied cavity modes $\ket{g_j,1_j}$ for $1 \leq j \leq j_M$, where $j_M$ is the number of emitter transitions to which the cavity couples. The states $\ket{g_j,1_j}$, henceforth known as `\eocs s', are each coupled to the excited state with respective coupling rate $g_j$. For atomic emitter-cavity applications using dipole-allowed transitions there may be three \eocs s for the three possible angular momentum transitions ($\pi$ or $\sigma_{\pm}$). However, the number of \eocs s is reduced if a decay is not allowed by atomic selection rules or if the cavity axis lies along the magnetic field and the $\pi$ decay is not supported, or increased if multiple state manifolds are close to resonance, for example due to hyperfine structure. Note that while the \eocs s $\ket{g_j,1_j}$ should be mutually orthogonal, neither the ground atomic states $\left\{\ket{g_j}\right\}$ nor occupied photon modes (whose single-occupancy states are denoted $\ket{1_j}$) need be mutually orthogonal. For non-birefringent cavities (where the two orthogonal polarisation modes associated with a spatial profile are degenerate), the most natural choice is to use a mutually orthogonal basis of atomic eigenstates $\left\{\ket{g_j}\right\}$, and define the corresponding photon states $\ket{1_j}$ through the cavity interaction. The level scheme is shown in Fig.~\ref{fig: multi level system} for an example of two \eocs s.  The equations of motion are

\begin{equation}
\begin{aligned}
\ket{\Psi(t)} & = \alpha_u \ket{u,0} + \alpha_e \ket{e,0} + \sum_{j=1}^{j_M}\alpha_{g_j}\ket{g_j,1_j}, \\
\dot{\alpha_u} & = -i\Delta_u\alpha_u - \Omega^*\alpha_e, \\
\dot{\alpha_e} & = -(\gamma+i\Delta_e)\alpha_e+\Omega\alpha_u+ \sum_{j=1}^{j_M}g_j\alpha_{g_j}, \\
\dot{\alpha_{g_j}} & = -(\kappa+i\Delta_{g_j})\alpha_{g_j}-g_j\alpha_e \,\,\, \forall j \, 1 \leq j \leq j_M, 
\end{aligned}
\label{eq: multi lambda equations of motion}
\end{equation}
where $\Delta_{g_j}$ is the detuning of \eocs{} $\ket{g_j,1_j}$ from an arbitrary reference level, which, for numerical convenience, is best chosen near the centre of the manifold of \eocs s.

The difficulty with optimising the outputs of this system is that, given appropriate boundary conditions, specifying an output wavefunction (for example~$\alpha_{g_1}(t)$) will intrinsically specify all other $\alpha_{g_j}(t)$ as occupied cavity-states couple only to one excited state $\ket{e,0}$. This interdependence is not straightforward to treat in the time domain as it involves both wavefunction terms and their time derivatives.

Instead, the wavefunction coefficients $\alpha_{g_j}(t)$ can be specified as a sum
\begin{equation}
\begin{aligned}
\alpha_{g_j}(t) & = \frac{1}{\sqrt{T_b}}\sum_n C_n^{(j)} e^{i\omega_n t}, \\
\omega_n & = \frac{2\pi}{T_b} n,
\end{aligned}
\label{double lambda system equations fourier}
\end{equation}	
of Fourier coefficients $C_n^{(j)}$ across a time domain of length $T_b$ which, as the photon wavefunctions start at zero amplitude but are not necessarily zero at $T$, must exceed $T$. The time-domain function $\alpha_{g_j}(t)$ can then be specified equivalently as a column vector in Fourier space,
\begin{equation}
\cvec{\alpha_{g_j}^F} = (..., C_{-1}^{(j)}, C_0^{(j)}, C_1^{(j)}, C_2^{(j)}, ...)^T.
\end{equation}
The equations of motion Eq.~(\ref{eq: multi lambda equations of motion}) relate the $\mathrm{n}^{\mathrm{th}}$ Fourier coefficient of the $\mathrm{k}^{\mathrm{th}}$ \eocs{} $\ket{g_k,1_k}$  to the same coefficient of the $\mathrm{j}^{\mathrm{th}}$ \eocs{} $\ket{g_j,1_j}$ through

\begin{equation}
C_n^{(k)} = \frac{g_k}{g_j}\frac{\kappa + i(\omega_n + \Delta_{g_j})}{\kappa + i(\omega_n + \Delta_{g_k})}C_n^{(j)} \equiv f_n^{(j\rightarrow k)} C_n^{(j)}.
\label{eq: double lambda system fourier conversion}
\end{equation}
Thus any output probability can be maximised with only the vector $\cvec{\alpha_{g_j}^F}$ as a variable because the wavefunction components in the remainder of the \eocs s may be automatically encoded in the relations of Eq.~(\ref{eq: double lambda system fourier conversion})

\subsection{Calculating Probabilities}

To maximise a desired output probability of the system, a selection of probabilities must be determined. These probabilities are
\begin{itemize}
\item $P_{\kappa_j}(t)$: The probability that a photon is emitted via $\ket{g_j,1_j}$ before time $t$.
\item $P_\gamma(t)$: The probability of spontaneous emission before time $t$.
\item $P_{g_j}(t)$: The probability that state $\ket{g_j,1_j}$ is occupied at time $t$.
\item $P_e(t)$: The probability that $\ket{e,0}$ is occupied at time $t$.
\end{itemize}
These probabilities can all be written as expectation values of matrices $\cmatrix{P_{\zeta}}$ with vector $\cvec{\alpha_{g_j}^F}$ such that
\begin{equation}
P_{\zeta}(t) = \cvec{\alpha_{g_1}^F}^\dag \cdot \cmatrix{P_{\zeta}}(t) \cdot \cvec{\alpha_{g_1}^F},
\label{eq: unnormalised general probability}
\end{equation}
for generic probability $P_{\zeta}$. Derivations of these matrices are presented in App.~\ref{app: calculating probabilities}, resulting in

\begin{equation}
\begin{aligned}
\cmatrix{P_{\kappa_j}}_{n',n}(t) & = 2\kappa(f_{n'}^{(1\rightarrow j)})^* \cmatrix{V}_{n', n}(t) f_n^{(1\rightarrow j)}, \\
\cmatrix{P_{g_j}}_{n',n}(t) & = (f_{n'}^{(1\rightarrow j)})^*(\exp \left\{i (\omega_n - \omega_{n'})t\right\})f_n^{(1\rightarrow j)}, \\
\cmatrix{P_{\gamma}}_{n',n}(t) & = \frac{2\gamma}{g_1^2}\left[\kappa^2+\Delta_{g_1}^2+ i\kappa(\omega_n-\omega_{n'})+\Delta_{g_1}(\omega_n + \omega_{n'})+\omega_n \omega_{n'}\right] \cmatrix{V}_{n', n}(t), \\
\cmatrix{P_e}_{n',n}(t) & = \frac{1}{g_1^2 T_b}\left[\kappa^2 + \Delta_{g_1}^2 +i\kappa\left(\omega_n-\omega_{n'}\right)+\Delta_{g_1}\left(\omega_n+\omega_{n'}\right) + \omega_n \omega_{n'}\right]\exp \left\{i (\omega_n - \omega_{n'})t\right\},
\end{aligned}
\end{equation}
where
\begin{equation}
    \cmatrix{V}_{n', n}(t) =
    \begin{drcases}
        \frac{2}{T_b\left(\omega_n-\omega_{n'}\right)} \sin\left(\frac{1}{2}\left(\omega_n-\omega_{n'}\right)t\right)\exp\left(\frac{1}{2}i\left(\omega_n-\omega_{n'}\right)t\right), & \omega_n \neq \omega_{n'}, \\
        \frac{t}{T_b}, & \omega_n = \omega_{n'}.
    \end{drcases}
\end{equation}

\subsection{Enforcing the Initial Vacancy of \eocs s}
\label{subsec: enforcing initial vacancy of eocs}
The solution must satisfy the constraint that there is no cavity occupation at $t=0$. Consider enforcing this condition on just the $j=1$ \eocs.
\begin{equation}
0 = \alpha_{g_1}(0) = \frac{1}{\sqrt{T_b}}\sum_n C_n^{(1)}.
\end{equation}
In the Fourier domain, this constraint is encoded
\begin{equation}
\begin{aligned}
\cvec{\phi_1^F}^{\dag} \cdot \cvec{\alpha_{g_1}^F} & = 0, \\
\left(\cvec{\phi_1^F}^{\dag}\right)_n & = 1.
\end{aligned}
\end{equation}
Equivalent conditions for the other \eocs s lead to a set of $j_M$ conditions,
\begin{equation}
\begin{aligned}
\cvec{\phi_j^F}^{\dag} \cdot \cvec{\alpha_{g_1}^F} & = 0 \, \,  \forall j, \\
\left(\cvec{\phi_j^F}^{\dag}\right)_n & = (f_n^{(1\rightarrow j)})^*.
\end{aligned}
\end{equation}
The $j_M$ vectors $\cvec{\phi_j^F}$ define a $j_M^d$-dimensional subspace in which a valid solution vector $\cvec{\alpha_{g_1}^F}$ should not lie, where $j_M^d \leq j_M$ is the number of non-degenerate \eocs s. We use Gram Schmidt orthogonalisation to produce a basis in which the final $j_M^d$ states span this subspace, along with a matrix $\cmatrix{U}$ that transforms a state in the Fourier basis to this basis. Additionally, we define the projector $\cmatrix{\Pi}_{j_M^d}$ which removes the last $j_M^d$ components of a vector, and the reverse projector $\tilde{\cmatrix{\Pi}}_{j_M^d}$ that takes a reduced vector and appends $j_M^d$ coefficients with value zero.
Any Fourier vector $\cvec{\alpha_{g_1}^F}$ can now be projected to a solution that satisfies the initial conditions
\begin{equation}
	\cvec{\alpha_{g_1}^P} = \cmatrix{\Pi}_{j_M^d}\cmatrix{U}\cvec{\alpha_{g_1}^F},
\end{equation}
where $\cvec{\alpha_{g_1}^P}$ is expressed as coefficients in the `projected' basis, to which the superscript $P$ refers. Probability matrices
\begin{equation}
\cmatrix{P_{\zeta}^P} =  \cmatrix{\Pi}_{j_M^d}\cmatrix{U}\cmatrix{P_{\zeta}}\cmatrix{U}^{\dag}\tilde{\cmatrix{\Pi}}_{j_M^d},
\end{equation}
are then written in the new projected basis where $\zeta$ is a generic index that specifies the probability, with these projected matrices again labelled by superscript $P$. Within the projected basis, every state automatically satisfies the initial conditions, and therefore these conditions need not be explicitly enforced during optimisation.

\subsection{Normalising Probabilities}
\label{subsec: normalising probabilities}
The (projected) matrix for total probability not in the initial state at time $t$
\begin{equation}
	\cmatrix{P_{\overline{u}}^P}(t) = \sum_{j=1}^{j_M} \left(\cmatrix{P_{\kappa_j}^P}(t)+\cmatrix{P_{g_j}^P}(t)\right)+\cmatrix{P_{\gamma}^P}(t)+\cmatrix{P_{e}^P}(t),
	\label{total probability definition double lambda}
\end{equation}
can be calculated from other probability matrices. As in the analysis of Sec.~\ref{sec: bounded approach}, a photon shape is possible if the total probability not in the initial state remains below unity for all times. We therefore define normalised probabilities as those found when the photon amplitude is re-scaled to the maximum that can satisfy this constraint. The normalised probability corresponding to a generic probability $P_{\zeta}(t)$ is
\begin{equation}
P^{N}_{\zeta}(t) = \frac{ \cvec{\alpha_{g_1}^P}^\dag \cdot \cmatrix{P_{\zeta}^P}(t) \cdot \cvec{\alpha_{g_1}^P}}{ \cvec{\alpha_{g_1}^P}^\dag \cdot \cmatrix{P_{\overline{u}}^P}(t_{\mathrm{max}}) \cdot \cvec{\alpha_{g_1}^P}},
\label{eq: auto-normalised probability}
\end{equation}
where $0\leq t_{\mathrm{max}}\leq T$ is the time for which $[\cvec{\alpha_{g_1}^P}^\dag \cdot \cmatrix{P_{\overline{u}}^P}(t_{\mathrm{max}}) \cdot \cvec{\alpha_{g_1}^P}]$ is maximised. This expression automatically normalises probabilities such that the sum of all calculated probabilities remains less than or equal to unity for the photon production process. Thus, when using normalised probabilities, the magnitude of the photon vector $\cvec{\alpha_{g_1}^P}$ has no significance.

\subsection{Optimising Probabilities}

To optimise probabilities, we use an iterative approach. Consider optimising a generic product of probabilities
\begin{equation}
V = \prod_{l=1}^{l_M} P^{N}_{\zeta_l}(t_l),
\label{non linear probability maximisation}
\end{equation}
where $l_M$ is the order of the probability product, $\zeta_l$ specifies the probability of the term in the product labelled by $l$, and $t_l$ is the time at which the probability should be evaluated. In general, a sum of such products can be desired, but this is just a trivial extension. 
Each cycle of an iterative procedure begins with the current solution vector $\cvec{\alpha_{g_1}^P}$. This vector is modified by adding a small correction

\begin{equation}
\delta \cvec{\alpha_{g_1}^P} = \epsilon \left[ \left( \sum_{l=1}^{l_M} \cmatrix{P_{\zeta_l}^P}(t_l) \cdot \cvec{\alpha_{g_1}^P}\right) -l_M \cmatrix{P_{\overline{u}}^P}(t_{\mathrm{max}}) \cdot \cvec{\alpha_{g_1}^P} \right],
\label{newton raphson delta phi}
\end{equation}
where $\epsilon$ is set to a randomly-chosen positive small number at each iteration to prevent the optimisation stalling and $t_{\mathrm{max}}$ is evaluated every iteration. This additional vector $\delta \cvec{\alpha_{g_1}^P}$ lies along the gradient of $V$ with respect to the projected solution vector $\cvec{\alpha_{g_1}^P}$ under the assumption that $\tmax$ does not change.

\subsection{Applicability to experiments}
The method presented in this section finds the output wavepacket that maximises a desired probability, however, this might not be the best practical solution for two main reasons. Firstly, physically realising the optimum output requires arbitrary control over the driving field, including effectively instantaneous transfers pulses at the start of, and potentially during, the process (the method to calculate the driving pulse from a solution for $\alpha_{g_1}(t)$ is given in App.~\ref{app: calculating driving}). In real experiments, the possible drive pulses are typically restricted, for example by the available drive power or modulator slew rates. It is much simpler to include these restrictions in optimal control approaches that optimise the driving parameters directly. Therefore, a sequential procedure first using our approach to find the performance limits fundamental to the emitter-cavity system followed by optimal control to find a realisable drive solution with performance close to these limits may be most effective. Secondly, the usefulness of photon wavepackets for a given application is often not solely a question of their total probability. For example, for applications involving photon interference, the wavepackets would ideally offer low sensitivity to inevitable pathlength differences \cite{Rohde:05} through reasonably stable intensity and phase profiles, a criterion which is not trivial to satisfy \cite{Morin:19}. Therefore a more general notion of photon desirability might be required. If these additional factors can be encoded as probability products, it may be possible to perform that optimisation within our framework, but again, a two-stage approach whose first step uses our method to understand the system limits may be more generally applicable. 

\subsection{Role of detunings}
Finally, it should be noted that, at no point during the optimisation, here or for the analytical approach in Sec.~\ref{sec: bounded approach}, did the detunings $\Delta_e$ and $\Delta_u$ feature. Therefore, the optimised wavepacket and probability are independent of their values. For $\Delta_u$, this is expected, because its value is only nominal as any change to its value could be compensated by the arbitrary drive pulse. However, $\Delta_e$ is a physical variable, so the conclusion that the optimum performance of these systems is unaffected by its value is not obvious, but this conclusion has been found in the adiabatic regime for photon generation~\cite{Goto:19} and storage~\cite{Giannelli:18}, and for the generation of Gaussian wavepackets~\cite{Utsugi:22}. As in these cases, the driving pulse to produce the optimum performance does depend on $\Delta_e$, even though the photon wavepacket and output probability does not.

\section{Results}
\label{sec: results}

In this section, we will use the analytic and numeric methods developed in Sections~\ref{sec: bounded approach} and \ref{sec: numerical approach} respectively to understand the limits of photon extraction in different scenarios. First, we will look at the $\Lambda$-system to evaluate the finite-time limits of photon extraction, the photon wavepackets that saturate these limits, and the scale of the benefits in decreased photon duration bestowed by these optimised wavepackets. Then, we focus on systems with more than one \eocs, demonstrating probability maximisation of different outputs, before performing a case study on how ground-state energy separation limits the success rates of systems designed for remote entanglement. Details on how numerical parameters such as the number of Fourier basis states were chosen are given in App.~\ref{app: selecting numerical parameters for simulations}.

\subsection{Optimum output of $\Lambda$-systems in finite time}

As mentioned in Sec.~\ref{subsec: three level model}, the optimum output from a $\Lambda$-system, $P_{\kappa}^{(a)}$ (Eq.~(\ref{eq: Lambda optimum output})), strictly applies only in infinite time. However, recent work studying Gaussian wavepackets~\cite{Utsugi:22}, in agreement with the numerical findings of~\cite{Giannelli:18}, indicates that output close to $P_{\kappa}^{(a)}$ is achieved provided the photon timescale ($T$ in our notation) is much greater than the `critical time'
\begin{equation}
    \tcrit = \mathrm{max}\left(\frac{\kappa}{g^2}, \frac{1}{\kappa}\right).
    \label{eq: critical time}
\end{equation}
This expression indicates two regimes for the ratio of $g$ to $\kappa$, with a boundary between them at $g=\kappa$.

An alternative to trying to drive a photon on timescale $T$ is to instantly excite the emitter to $\ket{e,0}$ at $t=0$ and wait for the excitation to transfer to the cavity mode and leak out to the collection. This technique produces photons quickly, but does so at an infinite-time probability
\begin{equation}
    P_{\kappa}^{(e)} = \frac{\kappa}{\kappa + \gamma} P_{\kappa}^{(a)},
    \label{eq: instant excitation probability}
\end{equation}
which is below the adiabatic limit~\cite{Cui:05}. The notions of $\tcrit$ and $P_{\kappa}^{(e)}$ highlight the trade-off that practical applications must balance between the rate of photon production $1/T$ and the extraction probability $P_{\kappa}(t)$. We use our analytic and numerical methods to find the limits of this trade-off.

\begin{figure}
\captionsetup{width=0.95\textwidth}
\centering 
\includegraphics*[width=1.0\linewidth]{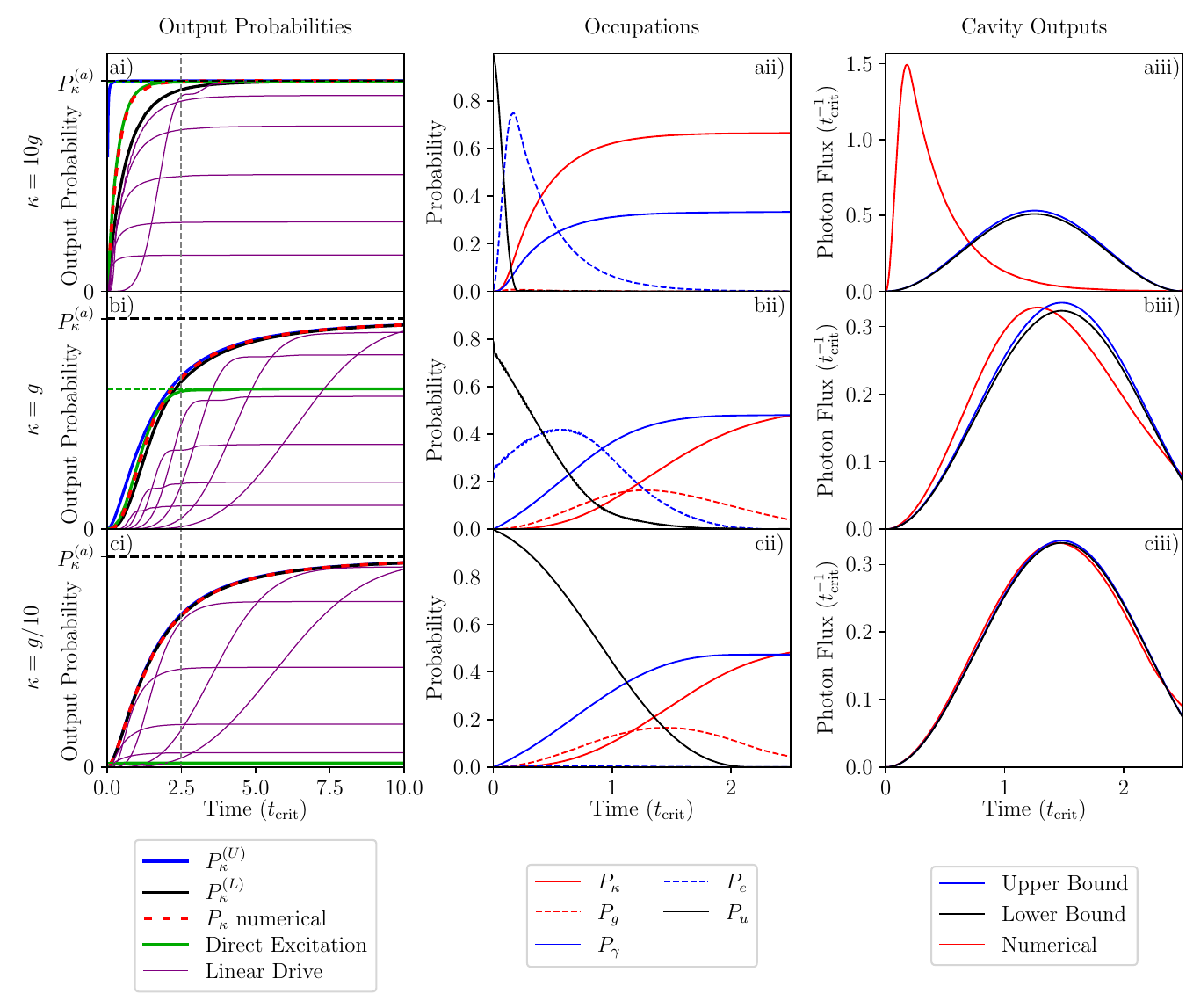}
\caption{Comparison of the numerically-optimised performance and analytical performance bounds of $\Lambda$-systems with $C=1$ and row a) $\kappa=10 g$, row b) $\kappa= g$, and row c) $\kappa=g/10 $. Column i) Output probabilities achieved through a range of methods against extraction time $T$. The blue and black lines mark the upper and lower bounds for the maximised extraction probability found in Sec.~\ref{sec: bounded approach} respectively. The red dashed line shows the numerically optimised output. The green line shows the extracted probability for direct excitation to the excited state, with the green dotted line marking the infinite-time extraction probability from direct excitation $P_{\kappa}^{(e)}$ (Eq.~(\ref{eq: instant excitation probability})). The purple lines show the output for a series of drives with linearly increasing amplitude and $\Delta_u=0$, where the rate of increase of drive amplitude is varied between different lines. The adiabatic extraction probability $P_{\kappa}^{(a)}$ (Eq.~(\ref{eq: Lambda optimum output})) is marked by the dotted black line, and takes a value of 2/3 for all scenarios. The example time used for analysing the system populations and photon wavepackets in ii) and iii) is $T = 2.5/\tcrit$, and is marked by the grey vertical line. Column ii) System populations for the numerically optimised solution for output time $T = 2.5/\tcrit$. Column iii) The probability flux of the output wavepackets for the numerically optimised case (red) compared to the upper (black) and lower (blue) bounds.} 
\label{fig: analytic_numeric_comparison}
\end{figure}

We examine these limits in Fig.~\ref{fig: analytic_numeric_comparison} for three example systems with cooperativity $C=1$, but with the ratio of $\kappa$ to $g$ chosen to sample the two regimes of $\tcrit$ (Eq.~(\ref{eq: critical time})) and the boundary between them. Examining first the comparison for different drive techniques over a range of times (Column i), the numerical result lies between the upper and lower bounds to the maximised photon extraction, which converge to $P_{\kappa}^{(a)}$ for $T \gg \tcrit$ as expected. The upper and lower bounds are quite distinct for the `bad cavity' (high $\kappa$, row a), but take similar values for $g\geq \kappa$. An instant excitation approach generally performs well at short times, but the output saturates at $P_{\kappa}^{(e)}$, which is particularly limiting for the parameters of row c). The linear drive curves provide an example of `simple' driving procedures, indicating how, particularly for the parameters of row a) and b), optimisation of the output wavepacket shape can lead to much faster photon production at a given efficiency. However, the performance of linear drives does tend to the maximum output probability in the adiabatic limit $T \gg \tcrit$, consistent with previous conclusions that photon extraction~\cite{Goto:19} and absorption~\cite{Gorshkov:07_1} probabilities reach this limit in infinite time regardless of wavepacket shape. 

Columns ii) and iii) show the system occupations and photon wavepackets respectively for case studies taking $T=2.5\tcrit$ respectively. Here we see that, in the case that $g \gg \kappa$ (row c), the optimised photon wavepacket has a very similar shape to the upper and lower bounds. However, as $\kappa$ increases (row b and then a), the wavepacket shape deviates significantly from the sinousoid form, instead exhibiting a sharp rise followed by a prolonged exponential decay.

\begin{figure}
\captionsetup{width=0.95\textwidth}
\centering 
\includegraphics*[width=1.0\linewidth]{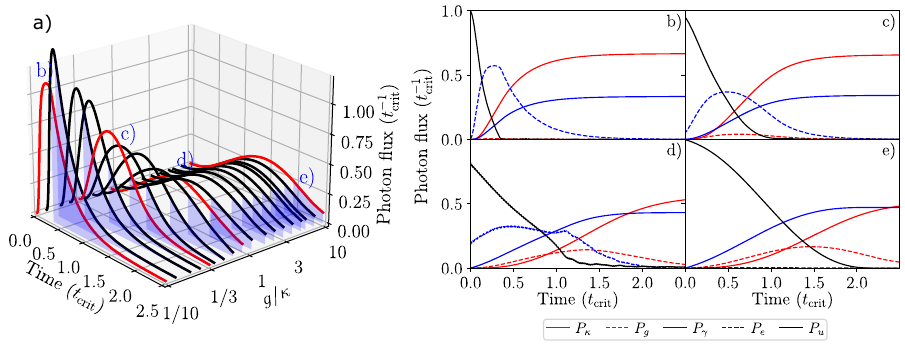}
\caption{Comparison of optimised cavity output profile as a function of $\Lambda$-system parameters. a) A collection of optimised output wavepackets found as the ratio of $g$ to $\kappa$ is varied, with $\gamma$ adjusted to maintain the cooperativity $C=1$, and the photon collection time set to $2.5 \tcrit$ for all data. The first time at which $P_u(t)$ is reduced below 1\% is marked by the beginning of the blue pane under each curve. Profiles taken as case studies are drawn with red lines, and are labelled in blue according to their corresponding probability panel. b)-e): System probabilities as a function of time for the corresponding output profiles labelled in a).} 
\label{fig: regime_sweep}
\end{figure}

To probe the variation in optimum wavepacket shape, the optimised wavepacket was calculated at $T = 2.5/\tcrit$ for a variety of $\kappa$:$g$ ratios, with the results shown in Fig.~\ref{fig: regime_sweep}. This shows that, in the limit $g \gg \kappa$, the optimised wavepackets tend to a single shape (when viewed in units normalised to $\tcrit$). This is because, for $g > \kappa$, Eq.~(\ref{eq: single lambda output final result}) always predicts the same upper bound in the normalised units, and the optimised wavepacket is very similar to this upper bound. However, as $\kappa/g$ increases, there is a gradual transition in shape towards the limit $\kappa \gg g$, which features a fast rise followed by a prolonged exponential decay. Profiles in the regime $\kappa > g$ also feature much more pronounced excited state occupations, which was also observed in~\cite{Utsugi:22}.

To understand why significant excited state population would distort the optimised wavepacket from the sinusoidal shape, it is important to remember that this shape was derived from the upper bound probability condition that $P_{\kappa}(T) + P_{\gamma}(T) + P_{g}(T) = 1$. The optimised wavepacket will only deviate from the upper bound shape if the true probability condition (that $P_{\kappa}(t) + P_{\gamma}(t) + P_{g}(t)$ + $P_e(t) = 1-P_{u}(t) = 1$ for some time $t$ during the process) is not equivalent to the upper bound condition. This requires either that $P_e(T) \neq 0$, or that $P_u(t)$ reaches zero at an intermediate time, which, given that $P_{\kappa}(t)$ and $P_{\gamma}(t)$ are monotonically increasing, would imply significant $P_e$ or $P_g$ at this intermediate time.

Observing the presented data again, we can broadly classify the observed profiles into two sets. For one set (Fig.~\ref{fig: regime_sweep}e), $P_u(t)$ tends smoothly towards zero at $T$ and the output approximates a sinusoidal profile. In the other set (Fig.~\ref{fig: regime_sweep}b, c, and d), $P_u(t)$ reaches zero at an intermediate time, and the output wavepacket features an exponential tail from this time onwards, indicated by the shape of the blue panes below the line in Fig.~\ref{fig: regime_sweep}a). Thus, for these cases, the probability condition is not a single restriction at the final time, but a continuous restriction for all later times. Therefore, in the $g \gg \kappa$ limit of the data, the dynamics are constrained by probabilities at the final time $T$, and are consequently dominated by the monotonically increasing probabilities $P_{\kappa}$ and $P_{\gamma}$. However, in the $\kappa \gg g$ limit, the dynamics are constrained by the probability sum condition over a period of times, meaning that the undulating probabilities $P_g$ and $P_e$ play an important role, and the shape derived in Sec.~\ref{sec: bounded approach} is not a good approximation. 

In summary then, the upper bound is easily calculable, but is not physically realisable, either because $P_e(T)$ is not exactly zero as assumed in the probability condition Eq.~(\ref{eq: upper bound probability constraint}), or the total probability not in the initial state would have to exceed unity at some point during the process. The lower bound is physically realisable, but it has an assumed, rather than optimum shape. The numeric method can find the wavepacket shape that optimises the output probability (which must lie between the bounds), however the wavepacket shape can be substantially different from that of the bounds. This optimum wavepacket shape can provide considerably reduced photon durations compared to simple linear drives operating at the same extraction efficiency, but the magnitude of this improvement depends strongly of the cavity parameter regime.

Finally, we note that the slight high-frequency noise seen in certain profiles (most obviously in Fig.~\ref{fig: analytic_numeric_comparison}bii) is likely a numerical artefact rather than a genuine feature. See App.~\ref{app: selecting numerical parameters for simulations} for more details.

\subsection{Optimisation of probability metrics}
\label{subsubsec: optimisation of probability metrics}
\begin{figure}
\captionsetup{width=0.95\textwidth}
\centering 
\includegraphics*[width=1.0\linewidth]{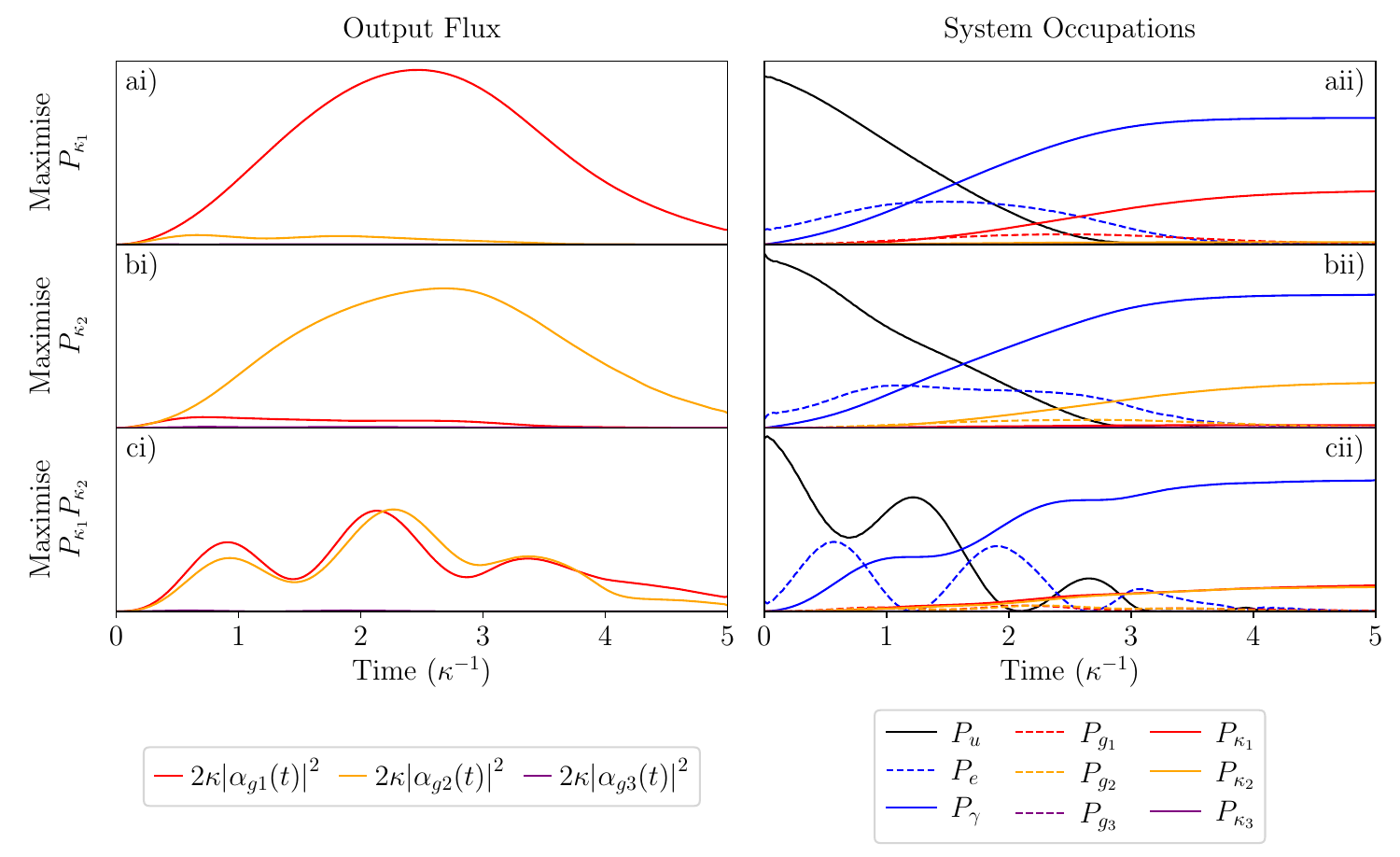}
\caption[Multi-level system targeted optimisations]{Example photon wavepackets (Column i) and system populations (Column ii) that maximise different probabilities for an emitter cavity system with $g_1 = \sqrt{1/3}\kappa$, $g_2 = -\sqrt{4/15}\kappa$, $g_3 = \sqrt{1/30}\kappa$, and $\gamma = 0.6\kappa$ over a time of $T=5/\kappa$. The energy splitting between $\ket{g_1,1_1}$ and $\ket{g_2,1_2}$, and $\ket{g_2,1_2}$ and $\ket{g_3,1_3}$, is $5\kappa$. Row a) shows the case where the maximised quantity is the single photon emission probability $P_{\kappa_1}$, row b) $P_{\kappa_2}$, and row c) the two-probability product $P_{\kappa_1}P_{\kappa_2}$.} 
\label{fig: optimised probabilities}
\end{figure}

The numeric method of Sec.~\ref{sec: numerical approach} is able to model emitter systems with multiple \eocs s, which is a common scenario for atomic emitters. Fig.~\ref{fig: optimised probabilities} shows how a single emitter-cavity system can be used to optimise different outputs. Here we take a system with $\gamma = 0.6\kappa$ and three \eocs s, such that $g_1 = \sqrt{1/3}\,\kappa$, $g_2 = -\sqrt{4/15}\,\kappa$, and $g_3 = \sqrt{1/30}\,\kappa$, which matches the example of an atomic $D_{\frac{5}{2}} \rightarrow P_{\frac{3}{2}}$ transition, where the initial state has angular momentum projection $m_J = 3/2$ along the magnetic field and the cavity is oriented orthogonal to the magnetic field. The energy splitting between \eocs s is $5\kappa$. The driving of this system was optimised for both strong single photon emission paths independently, $P_{\kappa_1}$ for row a) and $P_{\kappa_2}$ for row b), and the two-component decay probability $P_{\kappa_1}P_{\kappa_2}$, pertinent for protocols where emitter-photon correlations are desired, in row c). Note that this two-component probability product is not a two-photon event, but a maximisation of a product of two probabilities related to single photon emission. It should also be noted that, for the example cavity geometry and transitions, the photon-occupied states $\ket{1_1}$ and $\ket{1_3}$ would be identical (and $\ket{1_2}$ would have the orthogonal polarisation). This means that, while it is possible to distinguish $P_{\kappa_1}$ and $P_{\kappa_3}$ with an emitter measurement (justifying that they are distinct probabilities), it is not possible to distinguish them by photonic measurement. This must be understood when applying these methods to schemes where emitter measurement is not possible, for example remote entanglement schemes~\cite{Simon:03} where emitter measurement would destroy the intended entanglement.

The results show that the same cavity system can be used to produce relatively pure single photon emission through either strong decay channel, or a binary emitter-photon correlated state through changing the driving pulse. However, while the single probability-optimised scenarios (rows a and b) produce smooth wavepackets, the probability product-optimised wavepacket (row c) features significant oscillations. If the system were driven with the typical bichromatic drive~\cite{Stute:12_1}, then these oscillations would be attributed to off-resonant coupling between one drive tone and the `opposite' transition. However, our optimisation does not assume a specific form of driving pulse. We therefore see that it is not possible to produce non-undulating wavepackets without compromising on output probability. Finally, it should be noted that the method can optimise more general products or sums of probabilities, a faculty which may be used to discourage unwanted processes that contribute to error in a target application. 

\subsection{Case study: Effect of ground-state splitting on remote entanglement}
\label{subsec: case study remote entanglement}
Finally, we demonstrate the use of the numeric methods for understanding how emitter structure limits the performance of quantum protocols, taking two-photon, probabilistic, polarisation-encoded remote entanglement as a case study. In these schemes, two separate emitter-cavity systems are driven such that they both produce a single photon whose polarisation qubit is entangled with the final emitter state. For concreteness, we will imagine a system where $\ket{g_1}$ and $\ket{g_2}$ form the emitter qubit, and $\ket{1_1}$ and $\ket{1_2}$ have orthogonal polarisations so that the output wavepackets collectively constitute a polarisation qubit. The photon wavepackets from the two systems are routed to opposite input ports of a non-polarising beamsplitter and then subject to polarisation-resolved measurement. Upon certain outcomes of this measurement, the emitters are entangled (see~\cite{Simon:03, Luo:09} for more details of the scheme). Assuming two identical emitter-cavity systems are used, the probability of remote entanglement success in a given trial is proportional to $P_{\kappa_1}P_{\kappa_2}$~\cite{Moehring:07}. Even state-of-the-art free space~\cite{Nadlinger:22} and cavity-enhanced~\cite{Krutyanskiy:23} experimental implementations have success probability well below unity, leading to remote entanglement far slower than local operations, inhibiting the protocol's current usefulness. Given the importance of success probability and rate, it is crucial to understand exactly how different experimental parameters affect these quantities.

In an experimental scenario, the two pertinent \eocs s are not typically degenerate, with the splitting potentially chosen for reasons not directly related to the entanglement generation process. It is therefore of interest to know what impact a splitting (denoted $\Delta_Z$ as it would often constitute a Zeeman splitting in atom or ion applications) would have on the remote entanglement success probability.

\begin{figure}
\centering 
\captionsetup{width=0.95\textwidth}
\includegraphics*[width=1.0\linewidth]{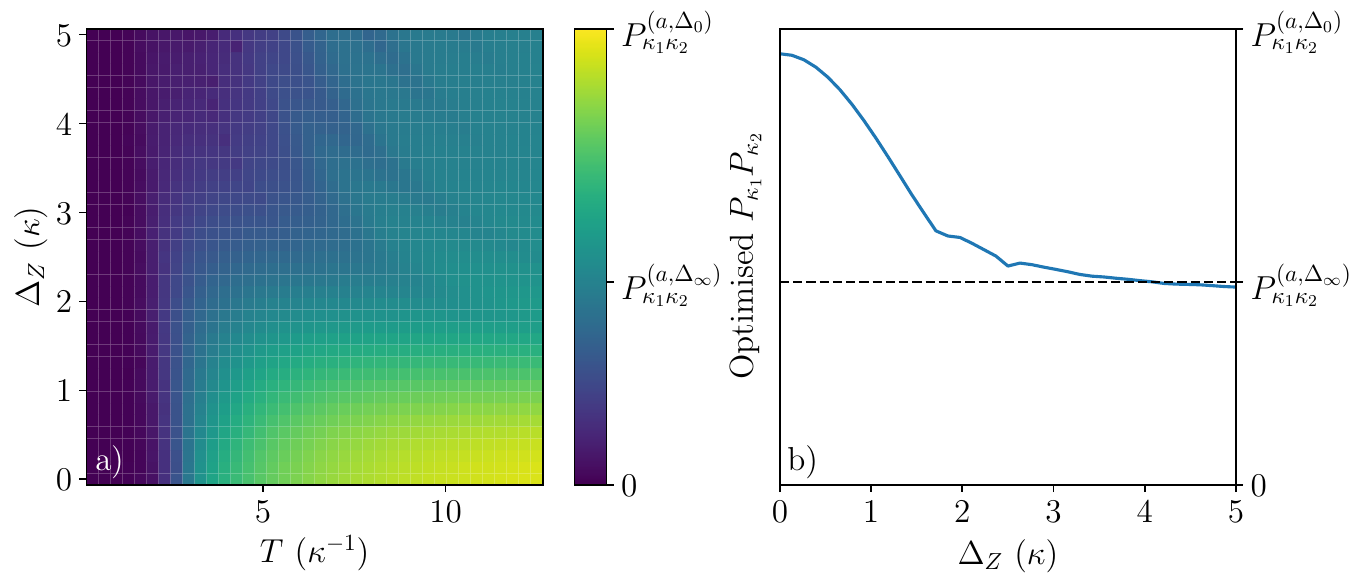}
\caption[Impact of Zeeman splitting on entanglement success probability]{Investigation of the effect of ground state splitting $\Delta_Z$ on the two-component probability $P_{\kappa_1}P_{\kappa_2}$ optimised for time $T$ for a system with $g_1 = \sqrt{1/3}\kappa$, $g_2 = -\sqrt{4/15}\kappa$ and $\gamma = 0.6\kappa$. a) The optimised $P_{\kappa_1}P_{\kappa_2}$ as a function of $T$ and $\Delta_Z$. b) The optimised output probability as a function of $\Delta_Z$ for just the longest time shown ($T=12.5/\kappa$) compared to the expected infinite time output for zero-splitting case $P_{\kappa_1 \kappa_2}^{(a, \Delta_0)}$ and for the high-splitting case $P_{\kappa_1 \kappa_2}^{(a, \Delta_{\infty})}$.} 
\label{fig: zeeman split map}
\end{figure}

We thus maximise $P_{\kappa_1}P_{\kappa_2}$ as a function of \eocs splitting $\Delta_Z$ and output time $T$ for the system investigated in Sec.~\ref{subsubsec: optimisation of probability metrics}, but where the third \eocs{} has been removed for simplicity. The results are shown in Fig.~\ref{fig: zeeman split map}, which shows broadly that an increasing \eocs{} splitting reduces the optimised remote entanglement success probability, with a high probability plateau for small $\Delta_Z$, a low plateau for large $\Delta_z$, and a transition in between.

Simple pictures for the limiting cases of photon extraction in limited time with $\Delta_Z=0$ and $\Delta_Z \rightarrow \infty$ can be used to understand the dependence of $P_{\kappa_1}P_{\kappa_2}$ on $\Delta_Z$, with derivations for these limits given in App.~\ref{app: limiting cases for remote entanglement study}. Firstly, in the case of $\Delta_Z=0$, the two decay channels have no differential phase evolution, so the excited state actually couples directly to a single level with rate $\effective{g}=\sqrt{g_1^2+g_2^2}$. This results in the infinite-time probability product
\begin{equation}
P_{\kappa_1 \kappa_2}^{(a, \Delta_0)} = \frac{g_1^2 g_2^2}{\effective{g}^4}\left(\frac{\effective{g}^2}{\effective{g}^2 + \kappa \gamma}\right)^2.
\label{eq: no splitting limit}
\end{equation}

Secondly, in the case that the Zeeman splitting becomes large, the production processes for the two components should be spectrally decoupled. This leads to an optimised $P_{\kappa_1}P_{\kappa_2}$ of
\begin{equation}
P_{\kappa_1 \kappa_2}^{(a, \Delta_{\infty})} = \frac{2C_1 2C_2}{4(2C_1+1)(2C_2+1)}.
\label{eq: well separated limit}
\end{equation}

The agreement of the data with Eqs.~(\ref{eq: no splitting limit}) and (\ref{eq: well separated limit}) is shown in panel b) of Fig.~\ref{fig: zeeman split map}. The data does not extend to infinite extraction time $T$, or to infinite state splitting $\Delta_Z$, but the trend broadly matches the values expected from these simple models.

While the conclusions of Fig.~\ref{fig: zeeman split map} are not themselves surprising, it is worth emphasising the key advantage of the optimisation method: that it is driving-independent. This means that we attribute the reduction in success probability with $\Delta_Z$ directly to the increase in $\Delta_Z$, without any doubt about whether we chose unsuitable driving pulses for some parameters. The same procedure could be applied to understand the dependence upon system parameters of the success probability of other quantum information protocols.

\section{Conclusions}
We have developed an analytic method to establish upper and lower bounds to the probability of photon extraction from a $\Lambda$-system in finite time which lie within the previously known infinite time bounds. We then extended these ideas to establish a numeric method to find the limits to photon extraction which also applies to generalisations of the $\Lambda$-system to include multiple cavity decays. This method can optimise general probability products, and, in-keeping with recent developments for related problems~\cite{Utsugi:22, Tissot:24}, does so without optimising the driving pulse explicitly. The combination of these approaches allows us to calculate the limits to the compromise between photon extraction probability and rate, and investigate optimised output wavepackets and corresponding quantum dynamics.

For the canonical $\Lambda$-system, we find the photon wavepacket which maximises finite-time photon extraction probability for given system parameters, observing that its shape lies on a spectrum between the exponential decay characteristic of fast excitation and the sinusoidal profile of the analytic bounds, and that this wavepacket can result in significantly faster high-efficiency photon extraction compared to simple driving approaches. In the case of generalised $\Lambda$-systems, which are often more accurate models of real experiments, we observed how to determine the limits of a single system, taken to model a trapped ion-cavity system, in producing single photons of specified polarisation, or an emitter photon correlated state. We then demonstrated how our methods can be used to derive driving-independent conclusions about the impact of system parameters, exemplified by the reduction of remote entanglement success probability with increasing \eocs{} splitting.

We believe that the methods presented in this manuscript will aid researchers in realising high-rate high-efficiency cavity-based sources for single photons and emitter-photon entangled states across the development cycle. Firstly, the methods determine the performance limits of existing hardware and how they can be reached. This specifically accounts for the finite protocol duration and extra emitter structure seen in experimental attempts to reach these limits~\cite{Schupp:21}. Secondly, the approach can reveal how the parameters of future systems will affect their performance, enabling the targeting of designs towards high performance use, which would again allow for the photon production time to be properly included in place of infinite time results~\cite{Gao:23}. Finally, the advantages of our driving-independent approach are complementary to that of optimal control, and the combination of the two may help identify robust driving schemes with near-ideal performance in practical systems. Additionally, we also reiterate that photon emission is deeply related to photon absorption, and our results in the context of photon emission experiments have direct analogues for the applications requiring finite-time absorption of photon wavepackets.

\section{Backmatter}
\noindent
\textbf{Funding}
This work was funded by the UK Engineering and Physical Sciences Research Council Hub in Quantum Computing and Simulation (EP/T001062/1), the UK Research and Innovation Frontier Research Fellowship (ERC Guarantee) MICRON-QC (EP/Y026438/1), and the European Union Quantum Technology Flagship Project AQTION (No. 820495). \\
\textbf{Acknowledgments}
The authors acknowledge the use of the IRIDIS High Performance Computing Facility, and associated support services at the University of Southampton, in the completion of this work. The authors would like to thank Dr. Thomas Doherty at the University of Oxford for careful reading of and insightful comments on this manuscript. \\
\textbf{Disclosures}
The authors declare no conflicts of interest. \\
\textbf{Data Availability Statement}
Data underlying the results presented in this paper are available in Ref.~\cite{dataset}. \\

\appendix


\section{Calculating probability functions for analytic method}
\label{app: calculating probability functions}
The probabilities used in Sec.~\ref{sec: bounded approach} of the main text can be expressed in terms of the integral notation of Eq.~(6), restated below for clarity
\begin{equation}
I_{(nm)}(t) = \int_0^t \left(\overset{n\cdot}{\alpha_g}\overset{m\cdot}{\alpha_g}^* + \overset{m\cdot}{\alpha_g}\overset{n\cdot}{\alpha_g}^*\right)\, dt,
\label{eq: integral notation definition appendix}
\end{equation}
where $n\cdot$ is a shorthand for $n$ copies of $\cdot$ indicating $n$ time derivatives.

Firstly, the probability $P_{\kappa}$ that there has been decay from the cavity mode is given by
\begin{equation}
\begin{aligned}
P_{\kappa}(t) & = 2\kappa\int_0^t \abs{\alpha_g(t')}^2 \, dt',\\
& = \kappa \int_0^t \alpha_g(t') \alpha_g^*(t') + \alpha_g^*(t') \alpha_g(t') \, dt',\\
&= \kappa I_{(00)}(t),
\end{aligned}
\end{equation}
where the final line follows from the definition of the notation in Eq.~(\ref{eq: integral notation definition appendix}).

The probability that the state is $\ket{g,1}$ can be written similarly through
\begin{equation}
\begin{aligned}
P_{g}(t) & = \abs{\alpha_g(t)}^2, \\
& = \int_0^t \frac{d}{dt'}\left(\abs{\alpha_g(t')}^2\right)\, dt', \\
& = \int_0^t \alpha_g(t')\dot{\alpha_g}^*(t') + \dot{\alpha_g}(t')\alpha_g^* (t') dt', \\
&= I_{(10)}(t),
\end{aligned}
\end{equation}
where the second line follows from the first because the state $\ket{g,1}$ is unoccupied at the start of the process.

The probability $P_{\gamma}(t)$ of spontaneous emission before time $t$ and $P_e(t)$ of occupation of state $\ket{e,0}$ at time $t$  are derived in a similar fashion, using Eq.~(\ref{eq: excited state wavefunction}) from the main text, which is restated below:
\begin{equation}
\alpha_e = -\frac{1}{g}\left(\kappa \alpha_g + \dot{\alpha_g} \right).
\label{eq: excited state wavefunction appendix}
\end{equation}
This relation may be substituted into the integral form for the spontaneous emission probability,
\begin{equation}
P_{\gamma}(t) = 2\gamma\int_0^t \abs{\alpha_e(t')}^2 \, dt',
\end{equation}
to yield
\begin{equation}
P_{\gamma}(t) = \frac{\gamma\kappa^2}{g^2}I_{(00)}(t) + \frac{2\gamma\kappa}{g^2}I_{(10)}(t) + \frac{\gamma}{g^2}I_{(11)}(t).
\end{equation}
Finally, the probability $P_e(t)$ that state $\ket{e,0}$ is occupied at time $t$ can be derived

\begin{equation}
\begin{aligned}
P_{e}(t) & = \abs{\alpha_e(t)}^2, \\
& = \int_0^t \frac{d}{dt'}\left(\abs{\alpha_e(t')}^2\right)\, dt', + \abs{\alpha_e(0)}^2 \\
& = \int_0^t \abs{-\frac{1}{g}\left(\kappa \alpha_g(t') + \dot{\alpha_g}(t') \right)}^2 + \abs{-\frac{1}{g}\left(\kappa \alpha_g(0) + \dot{\alpha_g}(0) \right)}^2 \\
&= \frac{\kappa^2}{g^2}I_{(10)}(t) + \frac{\kappa}{g^2}I_{(20)}(t) + \frac{\kappa}{g^2}I_{(11)}(t) + \frac{1}{g^2}I_{(21)}(t) + \frac{1}{g^2}\abs{\dot{\alpha_g}(0)}^2,
\end{aligned}
\end{equation}
where the third line follows from the second through the substitution of Eq.~(\ref{eq: excited state wavefunction appendix}), and the fourth from the third through the integral definitions Eq.~(\ref{eq: integral notation definition appendix}) and the initial condition $\alpha_g(0) = 0$.

The results of this appendix are summarised

\begin{equation}
\begin{aligned}
P_{\kappa}(t) &= \kappa I_{(00)}(t), \\
P_g(t) &= I_{(10)}(t), \\
P_{\gamma}(t) &= \frac{\gamma\kappa^2}{g^2}I_{(00)}(t) + \frac{2\gamma\kappa}{g^2}I_{(10)}(t) + \frac{\gamma}{g^2}I_{(11)}(t), \\
P_{e}(t) &= \frac{\kappa^2}{g^2}I_{(10)}(t) + \frac{\kappa}{g^2}I_{(20)}(t) + \frac{\kappa}{g^2}I_{(11)}(t) + \frac{1}{g^2}I_{(21)}(t) + \frac{1}{g^2}\abs{\dot{\alpha_g}(0)}^2, \\
P_{u}(t) &= 1 - P_{\kappa}(t) + P_g(t) +  P_{\gamma}(t) + P_{e}(t). 
\end{aligned}
\label{eq: simple lambda probabilities appendix}
\end{equation}

\section{The analytic upper bound does not have a hyperbolic solution}
\label{app: upper bound does not have hyperbolic solution}
The wavefunction amplitude $\alpha_g$ of the occupied cavity state for the upper bound of the emission probability satisfies Eq.~(\ref{eq: analytic alpha g eom}) of the main text, which is written below for clarity
\begin{equation}
\ddot{\alpha_g} + \kappa^2 \left(2 C q -1\right)\alpha_g = 0,
\label{eq: analytic alpha g eom appendix}
\end{equation}

where $\kappa$ is the cavity amplitude decay rate, $C$ is the cooperativity, and $q$ is a positive scalar. On the assumption that $2Cq - 1 < 0$, the solutions to Eq.~(\ref{eq: analytic alpha g eom appendix}) for $\alpha_g(t)$ are hyperbolic. The boundary condition that $\alpha_g(0) = 0$ means that the form must be a hyperbolic sine. Specifically, $\alpha_g$ satisfies

\begin{equation}
\begin{aligned}
\alpha_g & = A \sinh \left(s t\right), \\
s & = \sqrt{\kappa^2 \left(1 - 2 C q\right)},
\end{aligned}
\end{equation}
where $A$ is an undetermined amplitude. Substituting this wavefunction form into the equation for the system probabilities (Eq.~(\ref{eq: simple lambda probabilities}) from the main text) leads to
\begin{equation}
\begin{aligned}
P_{\kappa}(t) &= A^2 \kappa \left(\frac{1}{2s} \sinh\left(2 s t \right) - t \right), \\
P_g(t) &= A^2 \sinh^2 \left(s t\right), \\
P_{\gamma}(t) &= A^2 \left\{\frac{\gamma \kappa^2}{g^2}\left(\frac{1}{2s} \sinh\left(2 s t \right) - t \right) + \frac{2 \gamma \kappa}{g^2}\sinh^2 \left(s t\right) + \frac{\gamma s^2}{g^2}\left(\frac{1}{2s} \sinh\left(2 s t \right) + t \right)\right\}. \\
\label{eq: sinusoidal probabilities appendix}
\end{aligned}
\end{equation}

The second condition satisfied by the upper bound is that $q=m$ where $m=(P_{g}(T) + P_{\gamma}(T))/P_{\kappa}(T)$ (Eq.~(\ref{eq: single lambda minimisation setup}) from the main text). Evaluating $m$ with the probabilities from Eq.~(\ref{eq: sinusoidal probabilities appendix}) gives
\begin{equation}
m = \frac{1}{2C} + \frac{\left(\frac{2 \gamma}{g^2} + 1\right)\sinh^2 \left(s T\right)+ \frac{\gamma s^2}{\kappa g^2}\left(\frac{1}{2s} \sinh\left(2 s T \right) + T \right)}{\frac{1}{2s} \sinh\left(2 s T \right) - T}.
\end{equation}
The positivity of each term of the right hand side quotient for $T>0$ implies that $m \geq 1/2C$. This means that $q$ cannot equal $m$ and be consistent with our assumption that $2Cq - 1 < 0$. Therefore the hyperbolic case is never a solution for the upper bound wavepacket.

\section{Calculating probability matrices for the numeric method}
\label{app: calculating probabilities}

All probabilities $P_{\zeta}(t)$ used in the optimisation can be expressed as expectation values of matrices $\cmatrix{P_{\zeta}}(t)$ for Fourier vector $\cvec{\alpha_{g_1}^F}$. The simplest example is  $P_{\kappa_1}(t)$: 

\begin{equation}
\begin{aligned}
P_{\kappa_1}(t) & = 2\kappa \int_0^t \alpha_{g_1}^*(t')\alpha_{g_1}(t') \, dt', \\
& = \frac{2\kappa}{T_b} \int_0^t \sum_{n, n'}C_{n'}^{(1)*}C_n^{(1)} e^{i(\omega_n-\omega_{n'})t'} \, dt', \\
& = \sum_{n, n'} C_{n'}^{(1*)} (2\kappa \cmatrix{V}_{n', n}(t))C_n^{(1)}, \\
& = \cvec{\alpha_{g_1}^F}^\dag \cdot 2\kappa \cmatrix{V} \cdot \cvec{\alpha_{g_1}^F}, \\
\cmatrix{V}_{n', n}(t) & = \frac{1}{T_b}\int_0^t e^{i(\omega_n-\omega_{n'})t'} \, dt'. 
\end{aligned}
\label{PR1 calculation}
\end{equation}
This can be generalised to other emission probabilities by using the interdependence of the wavefunctions through
\begin{equation}
\begin{aligned}
P_{\kappa_j} & = \sum_{n, n'} C_{n'}^{(j)*} (2\kappa \cmatrix{V}_{n', n})C_n^{(j)}, \\
& = \cvec{\alpha_{g_1}^F}^\dag \cdot \left(2\kappa \cmatrix{F}^{(j)\dag}\cmatrix{V}(t) \cmatrix{F}^{(j)}\right) \cdot \cvec{\alpha_{g_1}^F}, \\
\cmatrix{F}_{n', n}^{(j)} & = f_n^{(1\rightarrow j)} \delta_{n',n}, 
\end{aligned}
\label{other PR calculations}
\end{equation}
where $\delta$ is the Kronecker delta, leading to the component form
\begin{equation}
    \cmatrix{P_{\kappa_j}}_{n',n}(t) = 2\kappa(f_{n'}^{(1\rightarrow j)})^* \cmatrix{V}_{n', n}(t) f_n^{(1\rightarrow j)}.
\end{equation}

For the cavity probabilities, the probability

\begin{equation}
\begin{aligned}
P_{g_j}(t) & = |\alpha_{g_j}(t)|^2, \\ 
& = \sum_{n, n'} C_{n'}^{(1)*} \left[(f_n^{(1\rightarrow j)})^*(\exp \left\{i (\omega_n - \omega_{n'})t\right\})f_n^{(1\rightarrow j)}\right]C_n^{(1)}, \\
& = \cvec{\alpha_{g_1}^F}^\dag \cdot \cmatrix{P_{g_j}} \cdot \cvec{\alpha_{g_1}^F}, \\
\cmatrix{P_{g_j}}_{n',n}(t) & = (f_{n'}^{(1\rightarrow j)})^*(\exp \left\{i (\omega_n - \omega_{n'})t\right\})f_n^{(1\rightarrow j)},
\end{aligned}
\label{PC1 calculation}
\end{equation}
is most easily expressed through a matrix given in component form.

The calculation for the spontaneous emission probability requires the expression for the excited state component (taken from Eq.~(\ref{eq: multi lambda equations of motion}) of the main text)
\begin{equation}
\alpha_e(t)= -\frac{\kappa+i\Delta_{g_1}}{g_1}\alpha_{g_1}-\frac{1}{g_1}\dot{\alpha_{g_1}}.
\end{equation}
From this, the spontaneous emission probability

\begin{equation}
\begin{aligned}
P_\gamma(t) & = 2\gamma \int_0^t \alpha_{e}^*(t')\alpha_{e}(t') \, dt', \\
& = \cvec{\alpha_{g_1}^F}^\dag \cdot \cmatrix{P_{\gamma}}(t) \cdot \cvec{\alpha_{g_1}^F}, \\
\cmatrix{P_{\gamma}}_{n',n}(t) & = \frac{2\gamma}{g_1^2}\left[\kappa^2+\Delta_{g_1}^2+ i\kappa(\omega_n-\omega_{n'})+\Delta_{g_1}(\omega_n + \omega_{n'})+\omega_n \omega_{n'}\right] \cmatrix{V}_{n', n}(t),
\end{aligned}
\label{PE calculation}
\end{equation}
can be expressed. The matrix for probability $P_e(t) = \abs{\alpha_e(t)}^2$ in the excited state at time $t$ can also be written in a similar component form 

\begin{equation}
\cmatrix{P_e}_{n',n}(t) = \frac{1}{g_1^2 T_b}\left[\kappa^2 + \Delta_{g_1}^2 +i\kappa\left(\omega_n-\omega_{n'}\right)+\Delta_{g_1}\left(\omega_n+\omega_{n'}\right) + \omega_n \omega_{n'}\right]\exp \left\{i (\omega_n - \omega_{n'})t\right\}.
\label{Pe definition}
\end{equation}
In order to calculate results numerically, the $\cmatrix{V}_{n', n}(t)$ can be evaluated exactly
\begin{equation}
    \cmatrix{V}_{n', n}(t) =
    \begin{drcases}
        \frac{2}{T_b\left(\omega_n-\omega_{n'}\right)} \sin\left(\frac{1}{2}\left(\omega_n-\omega_{n'}\right)t\right)\exp\left(\frac{1}{2}i\left(\omega_n-\omega_{n'}\right)t\right), & \omega_n \neq \omega_{n'}, \\
        \frac{t}{T_b}, & \omega_n = \omega_{n'}.
    \end{drcases}
\end{equation}

The matrices derived in this section can be summarised in component notation:
\begin{equation}
\begin{aligned}
\cmatrix{P_{\kappa_j}}_{n',n}(t) & = 2\kappa(f_{n'}^{(1\rightarrow j)})^* \cmatrix{V}_{n', n}(t) f_n^{(1\rightarrow j)}, \\
\cmatrix{P_{g_j}}_{n',n}(t) & = (f_{n'}^{(1\rightarrow j)})^*(\exp \left\{i (\omega_n - \omega_{n'})t\right\})f_n^{(1\rightarrow j)}, \\
\cmatrix{P_{\gamma}}_{n',n}(t) & = \frac{2\gamma}{g_1^2}\left[\kappa^2+\Delta_{g_1}^2+ i\kappa(\omega_n-\omega_{n'})+\Delta_{g_1}(\omega_n + \omega_{n'})+\omega_n \omega_{n'}\right] \cmatrix{V}_{n', n}(t), \\
\cmatrix{P_e}_{n',n}(t) & = \frac{1}{g_1^2 T_b}\left[\kappa^2 + \Delta_{g_1}^2 +i\kappa\left(\omega_n-\omega_{n'}\right)+\Delta_{g_1}\left(\omega_n+\omega_{n'}\right) + \omega_n \omega_{n'}\right]\exp \left\{i (\omega_n - \omega_{n'})t\right\}.
\end{aligned}
\end{equation}

\section{Calculating the required driving pulse}
\label{app: calculating driving}
Given a photon wavepacket $\alpha_{g_1}(t)$ for the numeric method (or equivalently $\alpha_g(t)$ for the analytic method), the driving to produce this wavepacket can be calculated. In practice, the driving should be calculated for $\alpha_{g_1}(t) = (1-\chi)\alpha_{g_1}^{S}(t)$, where $\alpha_{g_1}^{S}(t)$ is the ideal solution and $\chi$ is a small number. This is because the ideal solution is on the bounds of physical possibility, and therefore the driving required to produce it will tend to infinite strength at the end of the process~\cite{Vasilev:10}. Reducing the overall amplitude slightly removes this tendency. This is in analogy with an aspect of photon absorption where a photon wavepacket cannot be absorbed optimally by an emitter-cavity system without a singularity in the control drive at the initial time~\cite{Dilley:12}.  

Taking the equations of motion (Eq.~(\ref{eq: multi lambda equations of motion}) from the main text), all variables except the complex driving amplitude $\Omega$ and the obtained cavity output functions $\set{\alpha_{g_j}(t)}$ can be eliminated to produce a differential equation

\begin{equation}
\begin{aligned}
\dot{\Omega} & \left[\tilde{\kappa}\tilde{\gamma}\alpha_{g_1} +(\tilde{\kappa}+\tilde{\gamma})\dot{\alpha_{g_1}} + \ddot{\alpha_{g_1}} +g_1\sum_{j=1}^{j_M}(g_j\alpha_{g_j}) \right] = 
\\ & \phantom{+} \Omega [i\Delta_u \tilde{\kappa}\tilde{\gamma}\alpha_{g_1} + ( \tilde{\kappa}\tilde{\gamma}+ i\Delta_u(\tilde{\kappa}+\tilde{\gamma}))\dot{\alpha_{g_1}} \\
& \phantom{+ \Omega [}  + (\tilde{\kappa}+\tilde{\gamma}+i\Delta_u)\ddot{\alpha_{g_1}} + \dddot{\alpha_{g_1}} +g_1 \sum_{j=1}^{j_M}g_j(i\Delta_u  \alpha_{g_j}+\dot{\alpha_{g_j}})] \\
& + \abs{\Omega}^2 \Omega (\tilde{\kappa}\alpha_{g_1} + \dot{\alpha_{g_1}})
\end{aligned}
\label{eq: full system driving differential equation}
\end{equation}
where $\tilde{\gamma} \equiv \gamma + i\Delta_e$ and $\tilde{\kappa} \equiv \kappa + i\Delta_{g_1}$. These equations should be integrated from the initial driving amplitude

\begin{equation}
\begin{aligned}
\Omega^{(\theta_0)}(0, \theta_0) & = -\frac{(\tilde{\kappa}+\tilde{\gamma})\dot{\alpha_{g1}}(0)+\ddot{\alpha_{g1}}(0)}{g_1 \alpha_u^{(\theta_0)}(0)}, \\
\alpha_u^{(\theta_0)}(0) & = \sqrt{1-|\alpha_e(0)|^2}e^{i\theta_0}, \\
\alpha_e(0) & = -\frac{1}{g_1}\dot{\alpha_{g_1}}(0),
\end{aligned}
\label{eq: full system initial condition}
\end{equation}
where the superscript $\theta_0$ indicates that there is freedom in the initial condition. The shape of the cavity output may dictate that there is population in $\ket{e,0}$ at time $t=0$, meaning the initial state is a superposition of $\ket{u,0}$ and $\ket{e,0}$. The phase of this superposition is $\theta_0$, and does not affect whether the photon can be produced, but does affect the drive $\Omega$ that produces it. 

As discussed in Sec.~\ref{sec: bounded approach} of the main text, it is still possible to achieve the performance predicted from a system initialised to $\ket{u,0}$, but the driving now consists of two parts. The first is an instantaneous (or in practice very fast) excitation that sets up the initial state of Eq.~(\ref{eq: full system initial condition}), and the second is specified by Eq.~(\ref{eq: full system driving differential equation}). The relative phase of the two parts of the drive determines $\theta_0$, so must be precisely controlled.

It is important to note, however, that the conditions imposed have not restricted the driving in any way. As such, the drive may contain spikes of intensity where very fast changes in excited state amplitude are required to produce the optimum photon. These spikes can persist even when the target amplitude is reduced from the theoretical maximum amplitude.

The driving pulse derivative equation (Eq.~\ref{eq: full system driving differential equation}) is very sensitive to any numerical artefacts in the cavity output functions $\set{\alpha_{g_j}(t)}$. This means that the numerically optimised outputs from the main text can produce driving pulses that have very large swings in amplitude required to produce these artefacts. It is possible to reduce such issues by removing high frequency components from the optimised photon vector $\cvec{\alpha_{g_1}^F}$, but at the cost of distorting the wavepacket.

Once the full quantum dynamics have been calculated, the driving wavepacket can be verified by checking that it is consistent with a rearranged version of Eq.~(\ref{eq: multi lambda equations of motion}) from the main text
\begin{equation}
    \Omega = \left(\frac{1}{\alpha_u}\right)\left(\dot{\alpha_e} + \tilde{\gamma} \alpha_e - \sum_{j=1}^{j_M}g_j\alpha_{g_j} \right).
    \label{eq: driving verification}
\end{equation}
This equation could also be used to predict the driving in the $\Lambda$-system case with $\Delta_e = \Delta_u = \theta_0 = 0$. In that situation, the equations of motion (Eq.~(\ref{simple lambda system equations}) in the main text) are completely real, so wavefunction component amplitudes $\alpha_u$, $\alpha_e$, and $\alpha_g$ can be calculated directly from probabilities with the appropriate choice of sign. This special method for calculating the driving for particular $\Lambda$-systems is less sensitive to numerical artefacts than Eq.~(\ref{eq: full system driving differential equation}), but the impact of artefacts is still significant for our data. Therefore, to meaningfully calculate the driving for wavepackets optimised by the numeric procedure would likely require very smooth wavepackets that have been optimised extensively to remove any defects and very high time resolution.

\begin{figure}[htbp]
\centering
\includegraphics[width=.96\linewidth]{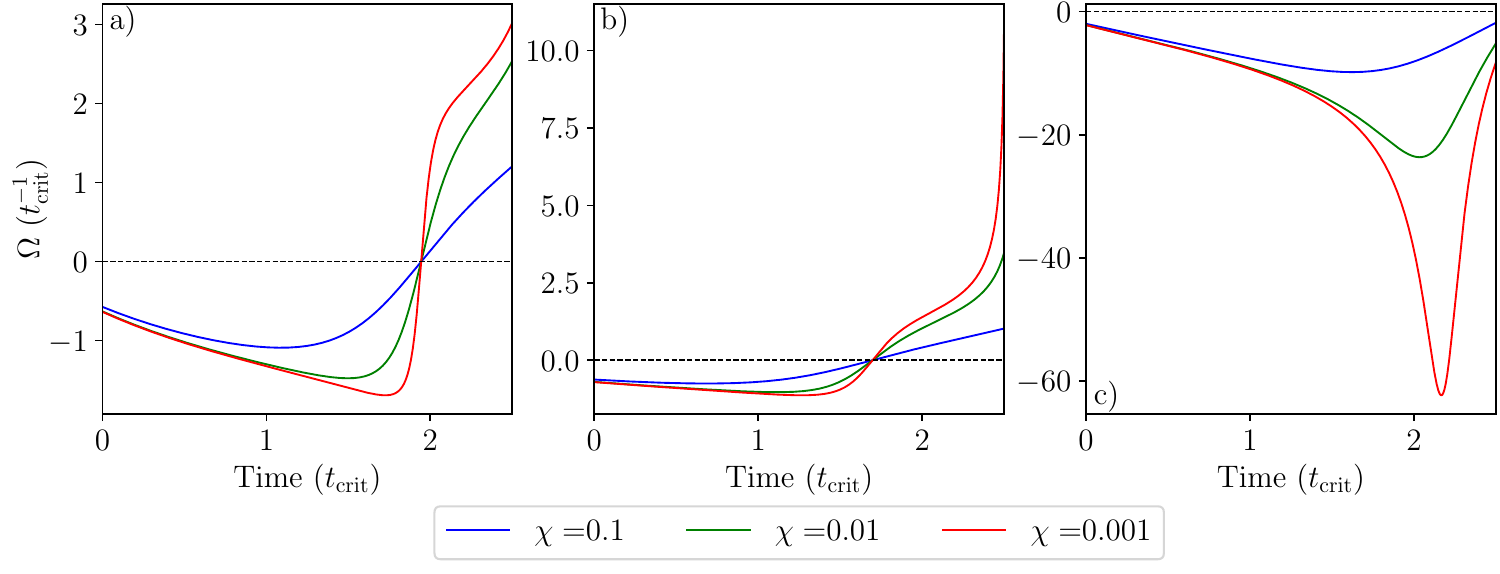}
\caption{Driving amplitude required to produce the lower bound wavepacket for the case studies presented in Fig.~\ref{fig: analytic_numeric_comparison} in the main text, in each case with $\Delta_e = \Delta_u = \theta_0 = 0$. Due to this choice, the driving amplitude $\Omega$ is always real in this figure, although for general parameters values $\Omega$ is a complex function of time. The plots a), b), and c) correspond to rows a), b), and c) in columns ii) and iii) of Fig.~\ref{fig: analytic_numeric_comparison} of the main text respectively. The driven photon wavepacket is scaled in amplitude by a factor $(1-\chi)$ compared to the true lower bound, meaning that $\lim_{\chi \to 0}$ produces the lower bound solution and is therefore on the edge of physical possibility.}
\label{fig:lower bound driving}
\end{figure}

To give a sense of the pulses required, it is possible to calculate the pulses using the lower bound wavepacket. This wavepacket has an analytically defined shape, and there are therefore no numerical artefacts. Examples of the driving required to produce the lower bounds for the case studies of Fig.~\ref{fig: analytic_numeric_comparison} in the main text were calculated using Eq.~\ref{eq: full system driving differential equation} and are shown in Fig.~\ref{fig:lower bound driving}. This shows how the driving strength required diverges as the probability limits are approached, and that the driving profile required varies considerably with system parameters. Finally, we see that the zeroes of the drive are a feature of the wavepacket shape and not its amplitude (as can be seen simply from Eq.~\ref{eq: driving verification}).

\section{Selecting numerical parameters for simulations}
\label{app: selecting numerical parameters for simulations}
\subsection{Number of Fourier basis states}
The numerical simulations optimise a desired probability by adjusting the vector $\cvec{\alpha_{g_1}^F}$ of amplitudes in a Fourier expansion of $\alpha_{g_1}(t)$. The number of Fourier coefficients used must be sufficient for the situation or the reported optimum will deviate from the true optimum. While there is no precise rule for the number of states that must be used, there are general principles to guide this choice. These can be formulated as conditions on the maximum basis frequency $\omega_{\mathrm{max}}$.

Firstly, the basis must be suitable large to describe the photon's temporal profile. This requires
\begin{equation}
    \omega_{\mathrm{max}} \gg \frac{2\pi}{\tau},
\end{equation}
where $\tau$ is the timescale of the photon. This timescale is at most $T$ (the total emission time of the photon), but for wavepackets with fast rises like the exponential decay solutions (Such as Fig.~3aiii), $\tau$ may be considerably shorter than $T$ and the requisite basis size correspondingly bigger.

Secondly, the wavepacket profile $\alpha_{g_1}(t)$ does not necessarily have to go to zero at time $T$, but, due to the initial condition at $t=0$ and the periodic temporal behaviour imposed by the Fourier series, must go to zero at the basis time $T_b$. This means that the basis must be sufficiently large to model the amplitude as it goes to zero between times $T$ and $T_b$, even if the amplitude in this time window is not physical. This imposes the condition
\begin{equation}
    \omega_{\mathrm{max}} \gg \frac{2\pi}{T_b - T}.
\end{equation}

Thirdly, with \eocs s detuned from the nominal zero frequency by $\Delta_{g_j}$, the Fourier expansion is likely to contain components at these frequencies. If the largest magnitude detuning in $\set{\Delta_{g_j}}$ is $\Delta_{\mathrm{max}}$, this imposes the condition
\begin{equation}
    \omega_{\mathrm{max}} > \Delta_{\mathrm{max}}.
    \label{eq: detuning frequency condition}
\end{equation}

Finally, to ensure that all \eocs s are initially unoccupied, a total of $j_M^d$ constraints are applied to the solution, where $j_M^d$ is the number of non-degenerate \eocs s (see Sec.~\ref{subsec: enforcing initial vacancy of eocs} in the main text). This means that the basis size must be reduced by $j_M^d$, and therefore must be bigger than $j_M^d$ to begin.

In practice the number of frequency states used for the investigation was typically around 50-100 positive frequencies (with an equal number of negative frequencies and a zero frequency component). However, this number varied significantly depending on the system parameters, with the wavepackets featuring prominent exponential decays generally requiring more components. An important limitation is imposed by the detuning condition Eq.~\ref{eq: detuning frequency condition} which means that a large number of basis states is required to model systems with large \eocs{} energy splittings and long photon production times.

\subsection{Number of normalisation times}
The method of ensuring that the probabilities during photon production do not violate probability conservation is described in Sec.~\ref{subsec: normalising probabilities} of the main text. This requires the time $t_{\mathrm{max}}$ at which the probability not contained in the initial state is maximised. To calculate this time, the unnormalised total probability matrix $\cmatrix{P_{\overline{u}}^P}(t)$ was calculated at a discrete series of times throughout the photon production process. At each step during the optimisation process of Eq.~(\ref{newton raphson delta phi}) in the main text, the maximum time $t_{\mathrm{max}}$ is calculated by finding the total probability matrix with the largest expectation.

A consequence of this method is that, if too few discrete times are chosen for evaluating the total probability, it is possible to violate probability conservation between discrete times, provided probability conservation is satisfied at discrete times. This is likely to be the reason behind the high frequency noise seen in some figures (such as \ref{fig: analytic_numeric_comparison}bii) or \ref{fig: regime_sweep}d) in the main text), which becomes stronger and lower frequency if the number of discrete times is reduced. The disadvantge to using a large number of discrete times is that the optimisation process takes considerably longer.

\section{Limiting cases for the remote entanglement case study}
\label{app: limiting cases for remote entanglement study}
For the remote entanglement case study (Sec.~\ref{subsec: case study remote entanglement} of the main text) we maximise the probability product $P_{\kappa_1}P_{\kappa_2}$ obtained in finite time from a generalised $\Lambda$-system with two \eocs s. The cavity coupling of one \eocs{} is $g_1$, and the other $g_2$, where the corresponding occupied cavity modes, $\ket{1_1}$ and $\ket{1_2}$, are orthogonally-polarised. The energy splitting $\Delta_Z$ between the two \eocs s ($\ket{g_1, 1_1}$ and $\ket{g_2, 1_2}$) and the time of photon production $T$ are varied. Simple models can be used to predict the infinite time outputs in the cases that $\Delta_Z=0$ and $\Delta_Z \rightarrow \infty$.

Firstly, in the case of $\Delta_Z=0$, the excited state actually couples directly to a single level 
\begin{equation}
\ket{g_{\mathrm{eff}}, 1_{\mathrm{eff}}} = \frac{g_1 \ket{g_1, 1_1} + g_2 \ket{g_2, 1_2}}{\sqrt{g_1^2+g_2^2}},
\end{equation}  
with an effective coupling $\effective{g}=\sqrt{g_1^2+g_2^2}$. This means that the adiabatic limit to the photon emission probability is

\begin{equation}
    \begin{aligned}
        P_{\kappa_{\mathrm{eff}}}^{(a, \Delta_0)} & = \frac{2 \effective{C}}{2 \effective{C} + 1}, \\
        \effective{C} & = \frac{\effective{g}^2}{2 \kappa \gamma}. 
    \end{aligned}
\end{equation}
The ratio of this emitted probability labelled $P_{\kappa_1}$ to the probability labelled $P_{\kappa_2}$ is $g_1^2:g_2^2$. The reported success probability in infinite time is therefore
\begin{equation}
P_{\kappa_1 \kappa_2}^{(a, \Delta_0)} = \frac{g_1^2 g_2^2}{\effective{g}^4}\left(P_{\kappa_{\mathrm{eff}}}^{(a, \Delta_0)}\right)^2,
\end{equation}
which may be rearranged to
\begin{equation}
P_{\kappa_1 \kappa_2}^{(a, \Delta_0)} = \frac{g_1^2 g_2^2}{\effective{g}^4}\left(\frac{\effective{g}^2}{\effective{g}^2 + \kappa \gamma}\right)^2.
\end{equation}

Secondly, in the case that $\Delta_Z$ becomes large, the production processes for the two components are spectrally decoupled. The decay indexed by $j=\{1,2\}$ then has a separate cooperativity $C_{j}=g_j^2/2\kappa\gamma$ and therefore an infinite-time ratio of cavity emission to spontaneous emission of $2C_j:1$. Note that this argument attributes some of the spontaneous emission probability $P_\gamma$ to \eocs{} 1 and some to \eocs{} 2. Though there is no experimental distinction between the spontaneous decay attributed to the two indices, this labelling functions to account for probabilities. If the total probability of either spontaneous emission attributed to decay channel 1 or emitted through \eocs{} 1 is $f_1$, the emitted probability through \eocs{} 1 is
\begin{equation}
    P_{\kappa_1} = f_1 \frac{2C_1}{2C_1 + 1},
\end{equation}
and equivalently for \eocs{} 2 the cavity emission probability is
\begin{equation}
    P_{\kappa_2} = f_2 \frac{2C_2}{2C_2 + 1},
\end{equation}
where $f_2$ is the sum of cavity emission and spontaneous emission attributed to \eocs{} 2. As there is no final occupation of the excited state or the \eocs s in the adiabatic regime, for an optimised output the sum of cavity and spontaneous emission outputs should be unity. This means that $f_1 + f_2 = 1$. It is simple to see that the optimum values of $f_1$ and $f_2$ are both $1/2$. This leads to an optimised success probability
\begin{equation}
P_{\kappa_1 \kappa_2}^{(a, \Delta_{\infty})} = \frac{2C_1 2C_2}{4(2C_1+1)(2C_2+1)}.
\label{eq: well separated limit appendix}
\end{equation}

\end{document}